%% file: Science_MainV3.tex
\documentclass[12pt]{article}

\usepackage[numbers,round,sort,longnamesfirst]{natbib}

\usepackage{setspace}

\usepackage{times}
\DeclareUnicodeCharacter{2061}{}

\usepackage{caption}
\usepackage{amsmath}
\usepackage{bm, mathtools}
\usepackage{accents} 
\usepackage{graphicx}
\graphicspath{../}

\usepackage{xcolor}
\usepackage{soul}
\usepackage{dcolumn}

\usepackage{float}
\usepackage{array}
\usepackage{booktabs}
\usepackage{multirow}

\usepackage{siunitx}
\usepackage[version=4]{mhchem}

\usepackage{etoolbox}

\def \Bz {k_{\rm{B}}}
\def \deg {^{\circ}}

\newcommand{\tcb}[1]{\textcolor{black}{#1}}

\newenvironment{sciabstract}{%
\begin{quote} \bf}
{\end{quote}}

\title{Universal Cold RNA Phase Transitions}

\author{
P. Rissone$^{*}$,$^{1}$ A. Severino$^{*}$,$^{1}$ I. Pastor,$^{1}$ F. Ritort$^{**}$.$^{1,2}$\\
\\
\normalsize{$^{1}$Small Biosystems Lab, Condensed Matter Physics Department,}\\ 
\normalsize{Universitat de Barcelona, C/ Marti i Franques 1, 08028 Barcelona, Spain}\\
\normalsize{$^{2}$Institut de Nanoci\`encia i Nanotecnologia (IN2UB),}\\
\normalsize{Universitat de Barcelona, 08028 Barcelona, Spain}\\
\\
\normalsize{$^{*}$These authors contributed equally to this work.}\\
\normalsize{$^{**}$To whom correspondence should be addressed; E-mail:  ritort@ub.edu}
}

\date{}

\begin{document} 


\baselineskip24pt


\maketitle 

\newpage

\begin{sciabstract}
    RNA's diversity of structures and functions impacts all life forms since {\it primordia}. We use calorimetric force spectroscopy to investigate RNA folding landscapes in previously unexplored low-temperature conditions. We find that Watson-Crick RNA hairpins, the most basic secondary structure elements, undergo a glass-like transition below $\mathbf{T_G\sim 20 ^{\circ}}$C where the heat capacity abruptly changes and the RNA folds into a diversity of misfolded structures. We hypothesize that an altered RNA biochemistry, determined by sequence-independent ribose-water interactions, outweighs sequence-dependent base pairing. The ubiquitous ribose-water interactions lead to universal RNA phase transitions below $\mathbf{T_G}$, such as maximum stability at $\mathbf{T_S\sim 5 ^{\circ}}$C where water density is maximum, and cold denaturation at $\mathbf{T_C\sim-50^{\circ}}$C. RNA cold biochemistry may have a profound impact on RNA function and evolution. 
\end{sciabstract}


\bigbreak
\noindent
Of similar chemical structure to DNA, the deoxyribose-ribose and thymine-to-uracil differences endow RNA with a rich phenomenology \cite{brion1997hierarchy, herschlag2018story, vicens2022thoughts}. RNA structures are stabilized by multiple interactions among nucleotides and water, often with the critical involvement of magnesium ions \cite{misra2002linkage, fiore2012entropic, yoon2014dynamical}. Such interactions compete in RNA folding, producing a rugged folding energy landscape (FEL) with many local minima \cite{chen2000rna,hyeon2003can}. To be functional, RNAs fold into a native structure via intermediates and kinetic traps that slow down folding \cite{russell2002exploring, woodson2010compact}. The roughness of RNA energy landscapes has been observed in ribozymes that exhibit conformational heterogeneity with functional interconverting structures \cite{woodson2000recent,xie2004single,solomatin2010multiple} and misfolding \cite{bassi1999rna,sinan2011azoarcus, bonilla2022cryo, li2022topological}. Single-molecule methods have revealed a powerful approach to investigate these questions by monitoring the behavior of individual RNAs one at a time, using fluorescence probes \cite{zhuang2000single,rueda2004single} and mechanical forces \cite{ritchie2015probing, bustamante2021optical}. Previous studies have underlined the crucial role of RNA-water interactions at subzero temperatures in liquid environments \cite{feig1998cryoenzymology,mikulecky2002cold,caliskan2006dynamic} raising the question of the role of water in a cold RNA biochemistry. Here, we carry out RNA pulling experiments at low temperatures, showing that fully complementary RNA hairpins unexpectedly misfold below a characteristic glass-like transition temperature $T_G\sim 20^{\circ}$C, adopting a diversity of compact folded structures. This phenomenon is observed in both monovalent and divalent salt conditions, indicating that magnesium-RNA binding is not essential for this to happen. Moreover, misfolding is not observed in DNA down to $5^{\circ}$C. These facts suggest that the folded RNA arrangements are stabilized by sequence-independent $2^{\prime}$-hydroxyl-water interactions that outweigh sequence-dependent base pairing. Cold RNA misfolding implies that the FEL is rugged with several minima that kinetically trap the RNA upon cooling, a characteristic feature of glassy matter \cite{kirkpatrick2015colloquium}. RNA folding in rugged energy landscapes is accompanied by a reduction of RNA's configurational entropy. A quantitative descriptor of this reduction is the folding heat capacity change at constant pressure, $\Delta C_p$, directly related to the change in the number of degrees of freedom available to the RNA molecule. Despite its importance, $\Delta C_p$ measurements in nucleic acids remain challenging \cite{chalikian1999more,mathews2002experimentally,mikulecky2006heat}. We carry out RNA pulling experiments at low temperatures and show that $\Delta C_p$ abruptly changes at $T_G\sim 20^{\circ}$C, a manifestation that the ubiquitous non-specific ribose-water interactions overtake the specific Watson-Crick base pairing at sufficiently low temperatures.


\bigbreak
\noindent
\textbf{\large RNA misfolds at low temperatures}

\begin{figure}[t]
\centering
\includegraphics[width=\textwidth]{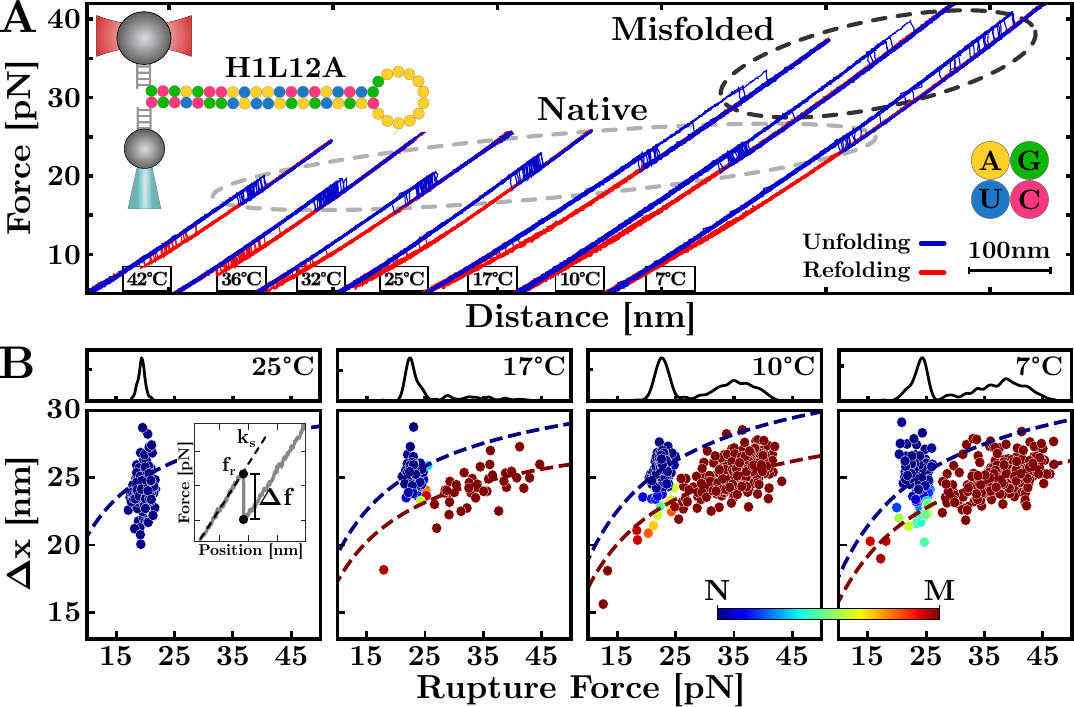}
\caption{\textbf{cold RNA misfolding.} (\textbf{A}) Unfolding (blue) and refolding (red) FDCs from H1L12A unzipping experiments (top-left) at temperatures $7-42^{\circ}$C and 4mM MgCl$_2$. The grey-dashed ellipse indicates native (N) unfolding events. Unexpected unfolding events from a misfolded (M) structure appear below $25^{\circ}$C (black-dashed ellipse) that become more frequent upon lowering $T$ from $17^{\circ}$C to $7^{\circ}$C. (\textbf{B}) Classification of N (blue dots) and M (red dots) rupture events at $T\leq 25^{\circ}$C and WLC fits for each state (dashed lines). The top panels show rupture force distributions at each $T$. The inset of the leftmost panel shows the parameters of rupture force events (see text).}
\label{fig:FIG1} 
\end{figure}

\noindent
We used a temperature-jump optical trap (\tcb{Sec. 1, Methods}) to unzip fully complementary Watson-Crick RNA hairpins featuring two 20bp stem sequences (H1 and H2) and loops of different sizes ($L=4,8,10,12$ nucleotides) and compositions (poly-A or poly-U) (\tcb{Sec. 2, Methods}). Pulling experiments were carried out in the temperature range $7-42^{\circ}$C at 4mM MgCl$_2$ and 1M NaCl in a 100mM Tris-HCl buffer (pH 8.1). Figure \ref{fig:FIG1}A shows the temperature-dependence of the force-distance curves (FDCs) for the dodeca-A (12nt) loop hairpin sequence H1L12A at 4mM magnesium. At and above room temperature ($T\geq 25^{\circ}$C), H1L12A unfolds at $\sim 20-25$pN (blue force rips in dashed grey ellipse), and the rupture force distribution is unimodal (Fig. \ref{fig:FIG1}B, leftmost top panel at $25^{\circ}$C), indicating a single folded native state (N). At $T \leq 17^{\circ}$C, new unfolding events appear at higher forces ($\sim 30-40$pN, dashed black ellipse). The bimodal rupture force distribution (Fig. \ref{fig:FIG1}B, right top panels) shows the formation of an alternative misfolded structure (M) that remains kinetically stable over the experimental timescales. Below $T = 10^{\circ}$C, the misfolded population shows $>50\%$ occupancy. 
Analogous results were obtained with sodium ions \tcb{(Fig. S2, Supp. Info)}. The formation of stable non-native structures for H1L12A is not predicted by algorithms such as Mfold \cite{zuker2003mfold}, Vienna package \cite{lorenz2011viennarna}, McGenus \cite{bon2013mcgenus}, pKiss \cite{janssen2015rna}, and Sfold \cite{ding2004s}. Furthermore, misfolding is not observed for the equivalent DNA hairpin sequence with deoxy-nucleotides \cite{rico2022temperature}. We refer to this phenomenon as cold RNA misfolding. 

Misfolding can be characterized by the size of the force rips at the unfolding events, which imply a change in the RNA molecular extension, $\Delta x$. The value of $\Delta x$ is obtained as the ratio between the force drop $\Delta f$ and the slope $k_{\rm s}$ of the FDC measured at the rupture force $f_r$, $\Delta x=\Delta f/k_{\rm s}$ (inset of left panel in Fig. \ref{fig:FIG1}B). Figure \ref{fig:FIG1}B shows $\Delta x$ versus $f_r$ for all rupture force events in H1L12A at four selected temperatures. Two clouds of points are visible below $25^{\circ}$C, evidencing two distinct folded states, the native (N, blue) and the misfolded (M, red). A Bayesian network model (\tcb{Sec. 3, Methods}) has been implemented to assign a probability of each data point belonging to N or M (color-graded bar in Fig. \ref{fig:FIG1}B). At a given force, the released number of nucleotides for N and M ($n_N$, $n_M$) is directly proportional to $\Delta x$ (\tcb{Sec. S2, Supp. Info}). To derive the values of $n_N$ and $n_M$, a model of the elastic response of the single-stranded RNA (ssRNA) is required. We have fitted the datasets ($\Delta x,f_r$) for N and M to the worm-like chain (WLC) elastic model (\tcb{Sec. 4, Methods}) using the Bayesian network model, finding $n_{N}=52(1)$ (blue dashed line) and $n_{M}=46(1)$ (red dashed line) for the number of released nucleotides upon unfolding the N and M structures. Notice that $n_N$ matches the total number of nucleotides in H1L12A (40 in the stem plus 12 in the loop), while M features 6nt less than $n_N$. These can be interpreted as remaining unpaired in M or that the $5^{\prime}-3^{\prime}$ end-to-end distance in M has increased by $\sim3$nm, roughly corresponding to 6nt.


\bigbreak
\noindent
\textbf{\large RNA flexibility at low-$T$ promotes misfolding}

\begin{figure}[t]
\centering
\includegraphics[width=\textwidth]{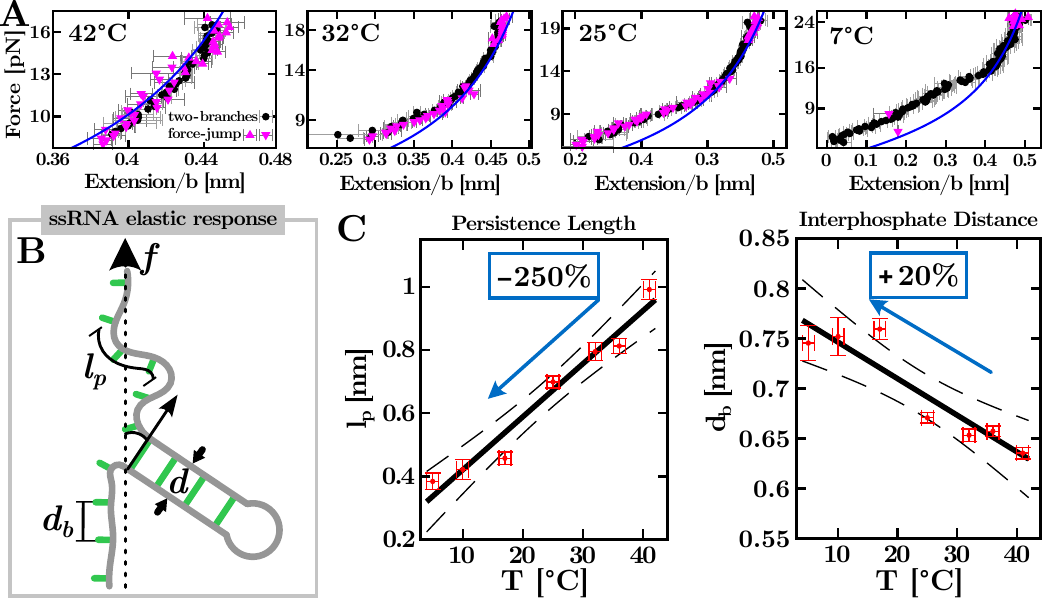}
\caption{\textbf{Temperature-dependent ssRNA elasticity}. (\textbf{A}) Force versus the ssRNA extension per base at different temperatures. Two methods have been used to extract the ssRNA molecular extension: the force-jump (magenta triangles up -- unfolding -- and down -- refolding --) and the two-branches method (black circles) \cite{zhang2009mechanoenzymatic,alemany2014determination}. Blue lines are the fits to the WLC in the high-force regime (see text). (\textbf{B}) Representation of the ssRNA elastic response according to the WLC model. The persistence length ($l_p$) measures the polymer flexibility, and the interphosphate distance ($d_b$) is the distance between contiguous bases. The computation of the total hairpin extension accounts for the contribution of the molecular diameter ($d$). (\textbf{C}) $T$-dependencies of $l_p$ (left) and $d_b$ (right). Linear fits (solid lines) with error limits (dashed lines) are also shown and give slopes equal to $0.17(2)$ \si{\angstrom}/K for $l_p$ and $-0.04(1)$ \si{\angstrom}/K for $d_b$.}
\label{fig:FIG2}
\end{figure}

\noindent
To characterize the ssRNA elasticity, we show the force-extension curves versus the normalized ssRNA extension per base in Fig. \ref{fig:FIG2}A for H1L12A at different temperatures. Upon decreasing $T$, the range of forces and extensions becomes wider due to the higher unfolding and lower refolding forces. Moreover, a shoulder in the force-extension curve is visible below $32^{\circ}$C (\tcb{see also Fig. S3, Supp. Info}), indicating the formation of non-specific secondary structures. A similar phenomenon has been observed in ssDNA \cite{viader2021cooperativity}. The force-extension curves (triangles and circles in Fig. \ref{fig:FIG2}A) at each temperature were fitted to the WLC model, with persistence length $l_p$ and inter-phosphate distance $d_b$ as fitting parameters (Fig. \ref{fig:FIG2}B and Eq.(1) in \tcb{Sec. 4, Methods}). Only data above the shoulder has been used to fit the WLC (\tcb{Sec. S1, Supp. Info}). The values $l_p$ and $d_b$ show a linear $T$-dependence (red symbols in Fig. \ref{fig:FIG2}C) that has been used for a simultaneous fit of the ssRNA elasticity at all temperatures (blue lines in Fig. \ref{fig:FIG2}A). Over the studied temperature range, $l_p$ (Fig. \ref{fig:FIG2}C, left panel) increases with $T$ by a factor of $\sim2.5$, whereas $d_b$ (Fig. \ref{fig:FIG2}C, right panel) decreases by only $\sim20\%$. The increase of $l_p$ with $T$ is an electrostatic effect \cite{rico2022temperature} that facilitates the bending of ssRNA at the lowest temperatures, promoting base contacts and misfolding. 


\bigbreak
\noindent
\textbf{\large Cold RNA misfolding is a universal sequence-independent phenomenon}

\begin{figure}[t]
\centering
\includegraphics[width=\textwidth]{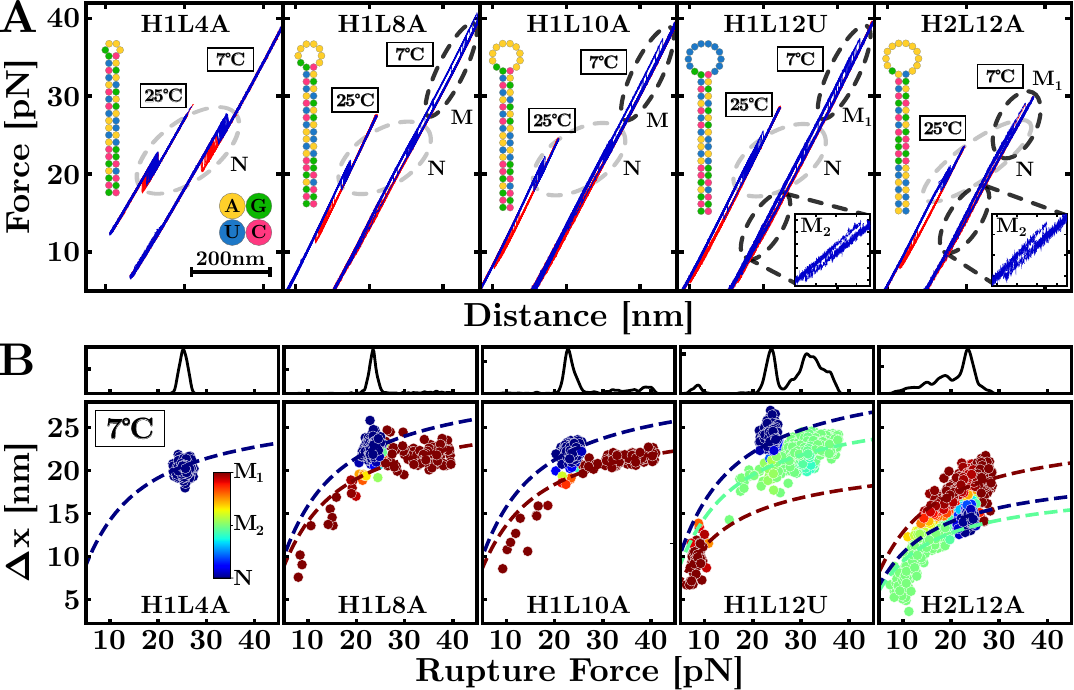}
\caption{\textbf{Universality of cold RNA misfolding.} (\textbf{A}) Unfolding (blue) and refolding (red) FDCs of hairpins H1L4A, H1L8A, H1L10A, H1L12U, and H2L12A at $25^{\circ}$C and $7^{\circ}$C. Grey-dashed ellipses indicate native (N) unfolding events. Except for H1L4A, all RNAs show unfolding events from misfolded (M) structures at $7^{\circ}$C (black-dashed ellipses). Hairpins H1L12U and H2L12A (featuring a dodeca-U loop and a different stem sequence) show a second misfolded structure at low forces (zoomed insets). Hairpin sequences are shown in each panel. (\textbf{B}) Bayesian classification of the unfolding events for the hairpins in panel (A) at $T=7^{\circ}$C. The dashed lines are the fits to the WLC for the different states. The top panels show the rupture force distributions.}
\label{fig:FIG3}
\end{figure}

\noindent
The ubiquity of cold misfolding is due to the flexibility of the ssRNA rather than structural features such as stem sequence, loop size, and composition. To demonstrate this, we show results for another five hairpin sequences in Fig. \ref{fig:FIG3}A with different stem sequences and loop sizes. To assess the effect of loop size, three hairpins have the same stem as H1L12A but tetra-A, octa-A, and deca-A loops (H1L4A, H1L8A, H1L10A respectively). A fourth hairpin features a dodeca-U loop (H1L12U) to avoid base stacking in the dodeca-A loop of H1L12A. The fifth hairpin, H2L12A, has the same loop as H1L12A but features a different stem. Except for H1L4A, all hairpins misfold below $T=25^{\circ}$C, as shown by the emergence of unfolding events at forces above 30pN (blue rips in the black dashed ellipses in Fig. \ref{fig:FIG3}A) compared to the lower forces of the unfolding native events $\sim 20$(grey dashed ellipses). Figure \ref{fig:FIG3}B shows the Bayesian-clustering classification of the different unfolding trajectories at $7^{\circ}$C and $25^{\circ}$C, in line with the results for H1L12A shown in Fig. \ref{fig:FIG1}B. The hairpin composition impacts misfolding; while H1L8A, H1L10A, and H1L12A show a single M at $7^{\circ}$C, H1L12U and H2L12A feature two distinct misfolded states at high (M$_1$) and low (M$_2$) forces (black dashed ellipses for H1L12U and H2L12A in Fig. \ref{fig:FIG3}A). 

\begin{figure}[t]
\centering
\includegraphics[width=\textwidth]{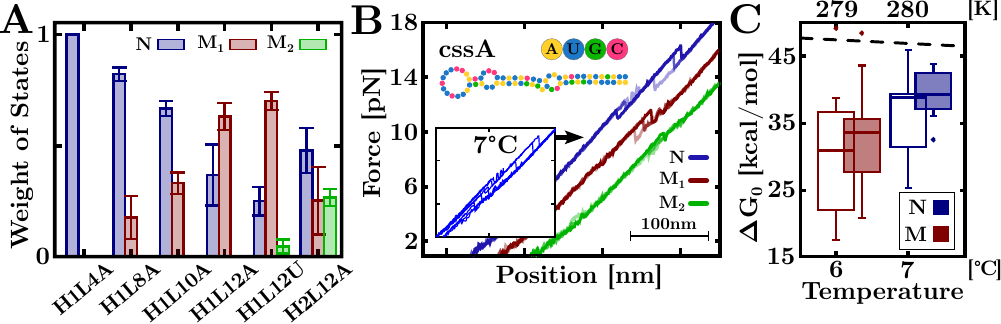}
\caption{\textbf{Features of cold RNA misfolding.} (\textbf{A}) Frequency of N, M$_1$, and M$_2$ unfolding events for the different RNA hairpins at $7^{\circ}$C. (\textbf{B}) Unfolding FDCs of cssA RNA at $7^{\circ}$C and 4mM \ce{MgCl2} (inset) classified into native (N) and misfolded (M$_1$, and M$_2$) states. (\textbf{C}) $\Delta G_0$ values at $7^{\circ}$C in 4mM \ce{MgCl2} (solid boxes) and 400mM \ce{NaCl} (empty boxes). Temperature axis in $^{\circ}$C (bottom label) and K (top label). Box-and-whisker plots show the median (horizontal thick line), first and third quartiles (box), 10th and 90th percentiles (whiskers), and outliers (dots). The black dashed line is the Mfold prediction.}
\label{fig:FIG4}
\end{figure}

The effect of the loop is to modulate the probability of formation of the native stem relative to other stable conformations. Indeed, H1L4A with a tetraloop has the largest stability among the studied RNAs \cite{varani1995exceptionally}, preventing misfolding down to $7^{\circ}$C (Fig. \ref{fig:FIG3}B). Misfolding prevalence increases with loop size due to the higher number of configurations and low entropic cost of bending the loop upon folding. The ssRNA elastic responses in H1L12A, H1L12U, and H2L12A show a systematic decrease of $l_p$ upon lowering $T$ (\tcb{Fig. S5, Supp. Info}) and therefore an enhancement of misfolding due to the large flexibility of the ssRNA.
Figure \ref{fig:FIG4}A shows the fraction of unfolding events at $7^{\circ}$C for all hairpin sequences for N (blue), M$_1$ (red), and M$_2$ (green). Starting from H1L4A, misfolding frequency increases with loop size, with the second misfolded state (M$_2$) being observed for H1L12U and H2L12A within the limits of our analysis. Compared to the poly-A loop hairpins (\tcb{Fig. S6, Supp. Info}), the unstacked bases of the poly-U loop in H1L12U confer a larger $d_b$ and extension to the ssRNA (red dots in \tcb{Fig. S5, Supp. Info}). \tcb{Elastic parameters for the family of dodecaloop hairpins are reported in Table S2, Supp. Info}. The fact that hairpins containing poly-A and poly-U dodecaloops misfold at low temperatures demonstrates that stacking effects in the loop are nonessential to misfolding. 

To further demonstrate the universality of cold RNA misfolding, we have pulled the mRNA of bacterial virulence protein CssA from \textit{N. meningitidis}, an RNA thermometer that changes conformation above 37$^\circ$C \cite{loh2013temperature}. Figure \ref{fig:FIG4}B shows several FDCs measured at $7^{\circ}$C and 4mM \ce{MgCl2} (inset), evidencing that the mRNA misfolds into two structures (M$_1$, red; M$_2$, green).


\bigbreak
\noindent
\textbf{\large RNA misfolds into stable and compact structures at low temperatures}

\noindent
The Bayesian analysis of the force rips has permitted us to classify the unfolding and refolding trajectories into two sets, $N\rightleftharpoons U$ and $M\rightleftharpoons U$ (Figs. \ref{fig:FIG1}B and \ref{fig:FIG3}B). We have applied the fluctuation theorem \cite{alemany2012experimental,rissone2022stem} to each set of trajectories of H1L12A to determine the free energies of formation of N and M from the irreversible work ($W$) measurements at $7^{\circ}$C \tcb{(Sec. 5, Methods and Sec. S4, Supp. Info)}.
In Fig. \ref{fig:FIG4}C, we show $\Delta G_0$ estimates for N (blue) and M (red), finding $\Delta G_0^N=38(9)$ kcal/mol and $\Delta G_0^M=30(10)$ kcal/mol in 4mM \ce{MgCl2} (empty boxes). We have also measured $\Delta G_0$ at 1M \ce{NaCl} and extrapolated it to 400mM \ce{NaCl}, the equivalent concentration to 4mM \ce{MgCl2} according to the 100:1 salt rule \cite{rissone2022stem}. We obtain $\Delta G_0^N=37(3)$ kcal/mol and $\Delta G_0^M=31(8)$ kcal/mol (filled boxes), in agreement with the magnesium data.
Within the experimental uncertainties, $\Delta G_0$ for N is higher by $\sim 5$ kcal/mol than for M, reflecting the higher stability of Watson-Crick base pairs in N. Notice that the Mfold prediction for N ($\Delta G_0^N=47$ kcal/mol, black dashed line) overestimates $\Delta G_0$ by 10 kcal/mol. 

We have also examined the distance between the folded and the transition state $x^{\ddag}$ in H1L12A to quantify the compactness of the folded structure. We have determined $x^{\ddag}$ from the rupture force variance $\sigma^2$ using the Bell-Evans model, through the relation $\sigma^2 = 0.61(k_{\rm B}T/x^{\ddag})^2$ (\tcb{Sec. S5, Supp. Info}). We find that average rupture forces for N and M decrease linearly with $T$, whereas $\sigma^2$ values are $T$-independent and considerably larger for M, $\sigma^2_{M}\sim 50 \sigma^{2}_{N}$, giving $x^{\ddag}_M=0.7(4)$nm and $x^{\ddag}_N=4.8(6)$nm (\tcb{Fig. S10, Supp. Info}). Therefore, $x^{\ddag}_M \ll x^{\ddag}_N$ with M featuring a shorter $x^{\ddag}$ and a more compact structure than N.


\bigbreak
\noindent
\textbf{\large The RNA glassy transition}

\begin{figure}[t]
\centering
\includegraphics[width=\textwidth]{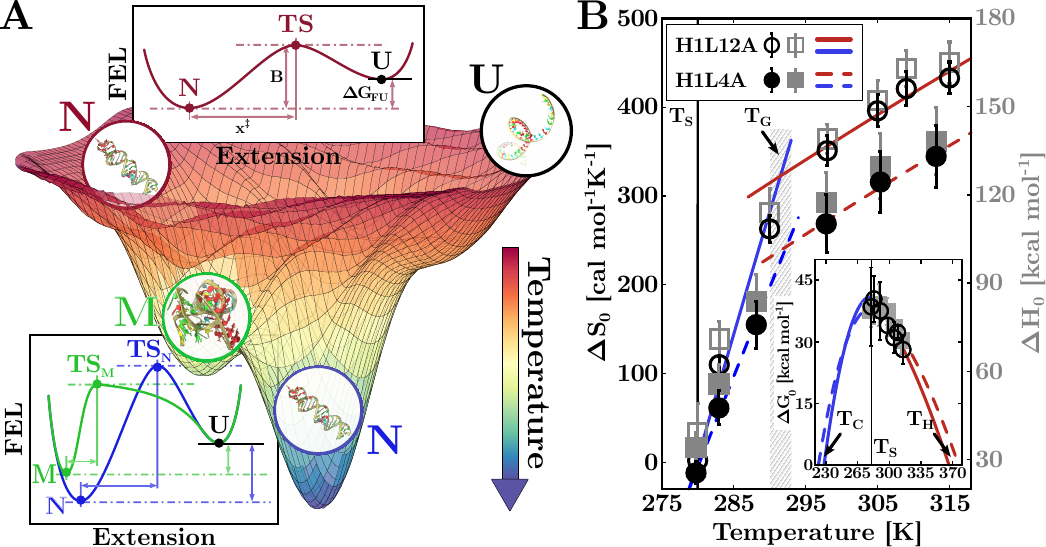}
\caption{\textbf{Cold RNA misfolding and phase transitions.} (\textbf{A}) Illustration of a multi-colored free-energy landscape (FEL) at different temperatures. The temperature arrow indicates the tendency to explore low-lying energy states with the FEL becoming rougher upon cooling: from high (red) to intermediate (green) and low (blue) temperatures. Transition state (TS) distances are typically shorter for M than N, denoting disordered and compact misfolded structures. The encircled schematic folds are for illustration purposes. (\textbf{B}) Temperature-dependent entropy (black) and enthalpy (grey) of N for H1L12A (empty symbols) and H1L4A (full symbols). The results are reported in \tcb{Table S5 and S6, Supp. Info}. Fits to the entropy values in the hot (red) and cold (blue) regimes for H1L12A (solid lines) and H1L4A (dashed lines) are also shown. The transition between the two regimes occurs at $T_G \rm \sim 293 K\sim 20^{\circ}$C (dashed grey band) with a sudden change in $\Delta C_p$. \textbf{Inset.} Stability curves of H1L12A (empty black circles) and H1L4A (solid grey squares). Maximum stability is found at $T_S \rm \sim 278K \sim 5^{\circ}$C (vertical black line) with melting temperatures at $T_H \rm \sim 370K \sim 100^{\circ}$C (red lines). Extrapolations of $\Delta G_0(T)$ in the cold regime predict cold denaturation at $T_C \rm \sim 220K\sim -50^{\circ}$C for both hairpins (blue lines).}
\label{fig:FIG5}
\end{figure}

\noindent
The ubiquity of the cold RNA misfolding phenomenon suggests that RNA experiences a glass transition below a characteristic temperature $T_G$ where the FEL develops multiple local minima. Figure \ref{fig:FIG5}A illustrates the effect of cooling on the FEL \cite{onuchic1997theory, gupta2011experimental}. Above $25^{\circ}$C, the FEL has a unique minimum for the native structure N (red-colored landscape). The projection of the FEL along the molecular extension coordinate shows that N is separated from U by a transition state (TS) (top inset, red line). Upon cooling, the FEL becomes rougher with deeper valleys, promoting misfolding (green and blue colored landscapes). The distance from M to TS is shorter than from N to TS, reflecting that M is a compact structure (bottom inset, green and blue lines). 

The glassy transition is accompanied by the sudden increase in the heat capacity change ($\Delta C_p$) between N and U below $T_G\sim 20^{\circ}$C for H1L12A and H1L4A. $\Delta C_p$ equals the temperature derivative of the folding enthalpy and entropy, $\Delta C_p=\partial \Delta H/\partial T=T\partial \Delta S/\partial T$ and can be determined from the slopes of $\Delta H(T)$ and $\Delta S(T)$ \tcb{(Sec. 6, Methods and Sec. S8, Supp. Info)}.
We observe two distinct regimes: above $T_G$ (hot, H) and below $T_G$ (cold, C). While $\Delta C_p^{\rm H} \sim 1.5\cdot 10^3$ cal mol$^{-1}$K$^{-1}$ is similar for both H1L12A and H1L4A (parallel red lines in Fig. \ref{fig:FIG5}B), $\Delta C_p^{\rm C}$ differs: $\Delta C_p^{\rm C}=8(1)\cdot 10^3$ cal mol$^{-1}$K$^{-1}$ for H1L12A versus $\Delta C_p^{\rm C}=5.8(4)\cdot 10^3$ cal mol$^{-1}$K$^{-1}$ for H1L4A (unparallel blue lines in Fig. \ref{fig:FIG5}B) showing the dependence of $\Delta C_p^{\rm C}$ on loop size at low-$T$. 
Despite the different $\Delta C_p^{\rm C}$ values, $\Delta S_0=0$ and stability ($\Delta G_0$) is maximum at $T_S=5(2)^{\circ}$C (Fig. \ref{fig:FIG5}B, inset) for both H1L4A and H1L12A (vertical black lines in Fig. \ref{fig:FIG5}B, main and inset). Finally, the $\Delta C_p^{\rm C}$ values predict cold denaturation at the same $T_C \sim -50^{\circ}$C for both sequences. 
The agreement between the values of $T_G$, $T_S$, and $T_C$ suggests that cold RNA phase transitions are sequence-independent, occurring in narrow and well-defined temperature ranges for all RNAs.


\clearpage

\bigbreak
\noindent 
\textbf{\large Discussion}

\noindent
Calorimetric force spectroscopy measurements on hairpin sequences of varying loop size, composition, and stem sequence show RNA misfolding at low-$T$ in monovalent and divalent salt conditions. The phenomenon's ubiquity leads us to hypothesize that non-specific ribose-water bridges overtake the preferential Watson-Crick base pairing of the native hairpin, forming compact structures at low temperatures. Cold misfolding is intrinsic to RNA, as it is not observed for the equivalent DNA hairpin sequences. In addition, magnesium ions are not crucial for it to happen, indicating the ancillary role of magnesium-mediated base-pairing interactions. Upon cooling, the diversity of RNA-water interactions promoted by the ribose increases the ruggedness of the folding energy landscape (FEL). Previous unzipping experiments of long (2kb) RNA hairpins at $25^{\circ}$C already identified stem-loops of $\sim 20$ nucleotides as the \textit{misfoldons} for RNA hybridization \cite{rissone2022stem}. The short RNA persistence length at low $T$ ($\sim 4\si{\angstrom}$ at $10^{\circ}$C, Fig. \ref{fig:FIG2}C) facilitates non-native contacts between distant bases and the exploration of different configurations. Indeed, the higher flexibility of the U-loop in H1L12U enhances bending fluctuations and misfolding compared to the stacked A-loop in H1L12A (Fig. \ref{fig:FIG4}A). Cold RNA misfolding has also been reported in NMR studies of the mRNA thermosensor that regulates the translation of the cold-shock protein CspA \cite{giuliodori2010cspa}, aligning with the CssA results of Fig. \ref{fig:FIG4}B. 
Cold RNA misfolding should not be specific to force-pulling but also present in temperature-quenching experiments where the initial high-entropy random coil state further facilitates non-native contacts \cite{hyeon2008multiple}. We foresee that cold RNA misfolding might help to identify \textit{misfoldon} motifs, contributing to developing rules for tertiary structure prediction \cite{laing2011computational,congzhou2023predicting}. 

Most remarkable is the large $\Delta C_p^{\rm C}$ values for H1L12A and H1L4A below $T_G\sim 20^{\circ}$C (293K), which are roughly 4-5 times the high-$T$ value above $T_G$, implying a large configurational entropy loss and a rougher FEL at low temperatures. 
The increase in $\Delta C_p$ below $T_G$ (dashed grey band in Fig. \ref{fig:FIG5}B) is reminiscent of the glass transition predicted by statistical models of RNA with quenched disorder \cite{pagnani2000glassy,iannelli2020cold}. As $\Delta C_p=C_p^U-C_p^N$, we attribute this change to the sudden reduction in $C_p^N$ and the configurational entropy loss upon forming N \cite{kirkpatrick2015colloquium}. Both hairpins show maximum stability $\Delta G_0$ at $T_S\sim 5^{\circ}$C (278K) where $\Delta S_0$ vanishes (Fig. \ref{fig:FIG5}B). The value of $T_S$ is close to the temperature where water density is maximum ($4^{\circ}$C), with low-$T$ extrapolations predicting cold denaturation at $T_C\sim -50^{\circ}$C (220K) for both sequences. This result agrees with neutron scattering measurements of the temperature at which the RNA vibrational motion arrests, $\sim 220$K \cite{caliskan2006dynamic, yoon2014dynamical}. We hypothesize that $T_S\sim 5^{\circ}$C and $T_C\sim -50^{\circ}$C mark the onset of universal phase transitions determined by the primary role of ribose-water interactions that are weakly modulated by RNA sequence, a result with implications for RNA condensates \cite{banani2017biomolecular, roden2021rna} and RNA catalysis \cite{wilson2021potential}. The non-specificity of ribose-water interactions should lead to a much richer ensemble of RNA structures and conformational states and more error-prone RNA replication. Cold RNA could be relevant for extremophilic organisms, such as psychrophiles, which thrive in subzero temperatures \cite{d2006psychrophilic}. Finally, misfolding into compact and kinetically stable structures might help preserve RNAs in confined liquid environments such as porous rocks and interstitial brines in the permafrost of the arctic soil and celestial bodies \cite{attwater2010ice, attwater2013ice}. This fact might have conferred an evolutionary advantage to RNA viruses for surviving during long periods \cite{wu2022permafrost} with implications on ecosystems due to the ongoing climate change \cite{cavicchioli2019scientists}. The ubiquitous sequence-independent ribose-water interactions at low temperatures frame a new paradigm for RNA self-assembly and catalysis in the cold. It is expected to impact RNA function profoundly, having potentially accelerated the evolution of a primordial RNA world \cite{higgs2015rna, joyce2018protocells}.




\bibliographystyle{Science}


\makeatletter
\apptocmd{\thebibliography}{\global\c@NAT@ctr 59\relax}{}{}
\makeatother

\section*{Supplementary References}

\renewcommand{\bibsection}{}


\section*{Acknowledgments}

We thank S. A. Woodson and C. Hyeon for a critical reading of the manuscript. 
\textbf{Funding}:
P.R. was supported by the Angelo Della Riccia Foundation. I. P. and F.R. were supported by Spanish Research Council Grant PID2019-111148GB-I00 and the Institució Catalana de Recerca i Estudis Avançats (F. R., Academia Prize 2018).
\textbf{Author contributions}: 
P.R., A.S., and I.P. carried out the experiments. P.R. and A.S. analyzed the data. I.P. and P.R. synthesized the molecules. I.P. and F.R. designed the research. P.R., A.S., and F.R. wrote the paper.
\textbf{Competing interests}:
The authors declare no competing financial interests.
\textbf{Data and materials availability}:
All data are available in the main text or the supplementary materials. This article has accompanying Supplementary Materials. 

\section*{Supplementary Materials}
    Materials and Methods\\
    Supplementary Text\\
    Supplementary Figs. S1 to S10\\
    Supplementary Tables S1 to S6\\

\clearpage


\newpage
\clearpage

\newpage

\newpage

\newpage

\newpage
%



\newpage
\include{Science_SuppInfoV3}



\end{document}

%% file: Science_SuppInfoV3.tex
\setcounter{page}{1}

\setstretch{1.0}  

\hspace{1cm}
\begin{center}
{\LARGE Supplementary Material for}\\
\vspace{0.4cm}
\textbf{\Large Universal Cold RNA Phase Transitions}\\
\hspace{0.5cm}

{P. Rissone$^{*}$, A. Severino$^{*}$, I. Pastor, F. Ritort.}

{\small $^{*}$Equally contributed authors.}

{\small Correspondence to: ritort@ub.edu}
\end{center}
\vspace{1cm}

\noindent

\noindent
\textbf{The PDF file includes:}
\begin{itemize}
    \setlength\itemsep{-0.4em}
    \item[] Materials and Methods
    \item[] Supplementary Text
    \item[] Supplementary Figs. S1 to S10
    \item[] Supplementary Tables S1 to S6 
\end{itemize}

\clearpage

{
\tableofcontents
}

\clearpage


\clearpage


\renewcommand{\thesection}{Material and Methods}
\renewcommand{\thesubsection}{\arabic{subsection}}

\section{}

\subsection{Temperature-jump optical trap}
\label{subsec:setup}

We used a temperature-jump optical trap to perform unzipping experiments at different temperatures \cite{de2015temperature}. Our setup adds to a MiniTweezers device \cite{rico2022molten} a heating laser of wavelength $\lambda = 1435$nm to change the temperature inside the microfluidics chamber. The latter is designed to damp convection effects caused by the laser non-uniform temperature, which may produce a hydrodynamics flow between medium regions (water) at different $T$. 
The heating laser allows for increasing the temperature by discrete amounts of $\Delta T \sim +2.5^{\circ}$C up to a maximum of $\sim +30^{\circ}$C with respect to the environment temperature, $T_0$. Operating the instrument in an icebox cooled down at a constant $T_0 \sim 5^{\circ}$C, and outside the box at ambient temperature (25$^{\circ}$C), we carried out experiments in the $T$ range $[7,42]^{\circ}$C. 

In a pulling experiment, the molecule is tethered between two polystyrene beads through specific interactions with the molecular ends \cite{bustamante2003ten}. One end is labeled with a digoxigenin (DIG) tail and binds with an anti-DIG coated bead (AD) of radius $3\mu$m. The other end is labeled with biotin (BIO) and binds with a streptavidin-coated bead (SA) of radius $2\mu$m.
The SA bead is immobilized by air suction at the tip of a glass micropipette, while the AD bead is optically trapped. The unfolding process is carried out by moving the optical trap between two fixed positions: the molecule starts in the folded state, and the trap-pipette distance ($\lambda$) is increased until the hairpin switches to the unfolded conformation. Then, the refolding protocol starts, and $\lambda$ is decreased until the molecule switches back to the folded state.

The unzipping experiments were performed at two different salt conditions: 4mM MgCl$_2$ (divalent salt) and 1M NaCl (monovalent salt). Both buffers have been prepared by adding the salt (divalent or monovalent) to a background of 100mM Tris-HCl (pH $8.1$) and $0.01\%$ NaN$_3$. The NaCl buffer also contains 1mM EDTA. The pulling protocols have been carried out at a constant pulling speed, $v=100$nm/s. We sampled 5-6 different molecules for each hairpin and at each temperature, collecting at least $\sim 200$ unfolding-folding trajectories per molecule.


\subsection{RNA synthesis}
\label{subsec:synthesis}

We synthesized six different RNA molecules made of a 20bp fully complementary Watson-Crick stem, ending with loops of different lengths ($L = 4, 8, 10, 12$ nucleotides) and compositions (poly-A or poly-U). The hairpins are flanked by long hybrid DNA/RNA handles ($\sim 500$bp). Further details about the sequences are given \tcb{Fig. S1 and Table S1, Supp. Info}. 

The RNA hairpins have been synthesized using the steps in Ref.\cite{bizarro2012non}. First, the DNA template (Merck, Township, NJ, USA) of the RNA is inserted into plasmid pBR322 (New England Biolabs, NEB, Ipswich, MA, USA) between the HindIII and EcoRI restriction sites and cloned into the \textit{E. coli} ultra-competent cells XL10-GOLD (Quickchange II XL site-directed mutagenesis kit). Second, the DNA template is amplified by PCR (KOD polymerase, Merck) using T7 promoters. The RNA is obtained by \textit{in-vitro} RNA transcription (T7 megascript, Merck) of the DNA containing the RNA sequence flanked by an extra 527 and 599 bases at the 3$^{\prime}$-end and 5$^{\prime}$-end, respectively, for the hybrid DNA-RNA handles. Finally, labeled biotin (5$^{\prime}$-end) and digoxigenin (3$^{\prime}$-end) DNA handles, complementary to the RNA handles, are hybridized to get the final construct.


\subsection{Bayesian clustering}
\label{subsec:clustering}

We use a mixture hierarchical Bayesian model (probabilistic graph network) to classify unfolding events as either emanating from a native or a misfolded initial folded state. The model is a soft classifier, giving each trace a probability (score) to belong to a given state. The model is described in \tcb{Sec. S3, Supp. Info}.


\subsection{ssRNA elastic model}
\label{subsec:ssRNAelastic}

The ssRNA elastic response has been modeled according to the worm-like chain (WLC), which reads
\begin{equation}
    \label{eq:WLC}
    f(x) = \frac{k_{\rm B}T}{4 l_p} \left[\left(1-\frac{x}{nd_b}\right)^{-2} - 1 + 4\frac{x}{nd_b} \right] \, ,
\end{equation}
where $l_p$ is the persistence length, $d_b$ is the interphosphate distance and $n$ is the number of bases of the ssRNA. More details on the WLC model and the fitting method used to derive its parameters can be found in \tcb{Sec. S1, Supp. Info}.


\subsection{Free energy determination}
\label{subsec:free_energy}

Given a molecular state, $\Delta G_0(N)$ is the hybridization free energy of the $N$ base pairs of the folded structure when no external force is applied ($f=0$). $\Delta G_0(N)$ is obtained from the free energy difference, $\Delta G_{\lambda}$, between a minimum ($\lambda_{\rm min}$) and a maximum ($\lambda_{\rm max}$) optical-trap positions where the molecule is folded and unfolded, respectively.
Thus, one can write
\begin{equation}
    \label{eq:dGtot}
    \Delta G(\lambda) = \Delta G_0(N) + \Delta G_{\rm el}(\lambda) \, ,
\end{equation}
where $\Delta G_{\rm el}(\lambda)$ is the elastic energy upon stretching the ssRNA between $\lambda_{\rm min}$ and $\lambda_{\rm max}$.
The latter term can be computed by integrating the WLC (Eq.\eqref{eq:WLC}). As unzipping experiments are performed by controlling the optical-trap position (not the force), this requires inverting Eq.\eqref{eq:WLC} (\tcb{Sec. S1, Supp. Info.}). 

We used the fluctuation theorem \cite{ciliberto2017experiments} (FT) to extract $\Delta G(\lambda)$ from irreversible work ($W$) measurements.
This is computed by integrating the FDC between $\lambda_{\rm min}$ and $\lambda_{\rm max}$, $W = \int_{\lambda_{\rm min}}^{\lambda_{\rm max}} fd\lambda$ (inset in \tcb{Fig. \ref{fig:H1L12AFT}}). 
Given the the forward ($P_{\rm F}(W)$) and reverse ($P_{\rm R}(W)$) work distributions, the FT reads
\begin{equation}
    \label{eq:FT}
    \frac{P_{\rm F}(W)}{P_{\rm R}(-W)} = \exp{\left( \frac{W - \Delta G(\lambda)}{k_{\rm B}T}\right)} \, ,
\end{equation}
where the minus sign of $P_{\rm R}(-W)$ is due to the fact that $W<0$ in the reverse process.
When the work distributions cross, i.e. $P_{\rm F}(W)=P_{\rm R}(-W)$, Eq.\eqref{eq:FT} gives $W = \Delta G(\lambda)$.
Let us notice that the FT can only be applied to obtain free-energy differences between states sampled under equilibrium conditions. However, 
pulling experiments at low $T$ are carried out under partial equilibrium conditions, with misfolding being a kinetic state. It is possible to extend the FT to our case by adding to $\Delta G(\lambda)$ from Eq.\eqref{eq:FT} the correction term $\Bz\log{(\phi^i_{\rm F}/\phi^i_{\rm R})}$, where $\phi_{\rm U(R)}$ is the fraction of forward (reverse) trajectories of state $i=\rm N, M$ \cite{alemany2012experimental}. 
Given $\Delta G(\lambda)$, the free energy at zero force, $\Delta G_0(N)$, is computed from Eq.\eqref{eq:dGtot} by subtracting the energy contributions of stretching the ssRNA, the hybrid DNA/RNA handles, and the bead in the trap. The first two terms are obtained by integrating the WLC in Eq.\eqref{eq:WLC}), while the latter is modeled as a Hookean spring of energy $\Delta G_b(x) = 1/2 k_{\rm b} x^2$, where $k_{\rm b}$ is the stiffness of the optical trap.


\subsection{Derivation of the heat capacity change}
\label{subsec:heatcapacity}

To derive $\Delta C_p$, we have measured the enthalpy $\Delta H_0$ and entropy $\Delta S_0$ of N at different $T$'s for H1L12A and H1L4A. $\Delta S_0$ is obtained from the extended form of the Clausius-Clapeyron equation in a force \cite{rico2022molten}, while $\Delta H_0=\Delta G_0+T\Delta S_0$. Both $\Delta H_0$ and $\Delta S_0$ are temperature dependent, with a finite $\Delta C_p$ (\tcb{Sec. \ref{subsec:dCp}, Supp. Info}). This has been obtained by fitting the $T$-dependent entropies to the thermodynamic relation $\Delta S_0(T) = \Delta S_m + \Delta C_p\log{(T/T_m)}$, where $T_m$ is the reference temperature and $\Delta S_m$ is the entropy at $T=T_m$.

\clearpage


\renewcommand{\thesection}{Supplementary Text}
\section{}

\renewcommand{\thefigure}{S\arabic{figure}}
\renewcommand\theequation{S\arabic{equation}}
\renewcommand{\thesection}{S\arabic{section}}
\renewcommand{\thetable}{S\arabic{table}}
\setcounter{section}{0}
\setcounter{figure}{0}
\setcounter{equation}{0}


\section{Worm-like chain model}
\label{sec:WLC-SI}

\subsection{Explicit inversion}
\label{subsec:WLCinv}

We describe the ssRNA elastic response according to the worm-like chain (WLC) model \cite{bustamante1994entropic, marko1995stretching}, which reads
\begin{equation}
    f_{\rm WLC}(x) = \frac{k_{\rm B}T}{4 l_p} \left[\left(1-\frac{x}{nd_b}\right)^{-2} - 1 + 4\frac{x}{nd_b} \right] \, ,
\end{equation}
where $l_p$, $d_b$, and $n$ are the persistence length, interphosphate distance, and number of monomers in the ssRNA, respectively. 
Note that the WLC model expresses the force as a function of the extension and can be inverted \cite{severino2019efficient} to obtain the extension per monomer $z \equiv x / n d_b$ as a function of the force:
\begin{equation} \label{eq:inv-WLC}
    z(f) = f_{\rm WLC}^{-1}(f) \, .
\end{equation}
The inverted form of the WLC, $z(f)$, speeds up data analysis and is implemented in the JAGS library for the Bayesian classification (\tcb{Sec. \ref{subsec:Bayesian-clustering}, Methods}).


\subsection{Multi-$T$ elastic response fit}
\label{subsec:multiT_fit}

We fit the temperature dependence of the elastic parameters ($l_p$, $d_b$) to the WLC model, assuming they are linearly dependent on temperature. The assumption is supported by available experimental evidence (\tcb{Fig. 2C, main text}). Therefore, we use the fitting expressions:
\begin{subequations}
\label{eq:multiFIT}
    \begin{align}
    	\label{eq:lp_T}
    	l_p = l_p(T) &= a_1 T + b_1 \\
    	\label{eq:db_T}
    	d_b = d_b(T) &= a_2 T + b_2 \, .
    \end{align}
\end{subequations}

The result of this multi-$T$ fitting procedure is shown in \tcb{Fig. \ref{fig:ElasticMultiFit}, Supp. Info}. Notice that to avoid the secondary structure plateau (which cannot be described by the WLC model), only data points in the high force range of the elastic response above the shoulder in the data have been used.


\section{Released nucleotides in unfolding events} 
\label{subsec:number-nt}

In an unfolding event, the extension of the ssRNA released in the transition from the folded to the unfolded state, $x_r(f)$, can be obtained from the experimental FDCs through the relation
\begin{equation}
    \label{eq:x-r}
    x_r(f^U) = \frac{\Delta f}{k_{\mathrm{eff}}^{F}} + x_d (f^U),
\end{equation}
where $\Delta f = f^F - f^U$ is the force difference upon unzipping between the force in the folded branch $F$ ($f^F$) and in the unfolded branch $U$ ($f^U$), $k_{\mathrm{eff}}^{F}$ is the effective stiffness in the folded branch, i.e. the slope of the FDC before unfolding, and $x_d$ is the diameter of the folded structure projected along the pulling axis.
We used the value of $x_r$ determined from Eq.\eqref{eq:x-r} to asses whether the unfolding events experimentally observed originate from the native state (hairpin) or misfolded state. Given the number of nucleotides in the folded structure, $n$, the following relation holds:
\begin{equation} \label{eq:bayes-precursor}
    x_r(f^U) = n \cdot d_b \cdot f_{\rm WLC}^{-1}(f^U) \, .
\end{equation}
Therefore, different states characterized by different $n$ give different ($f^U$, $x_r(f^U)$) distributions, as shown in \tcb{Fig. 1B and 3B, of the main text}. 

The advantage of Eq.\eqref{eq:bayes-precursor} is two-fold. First, by assuming that the WLC parameters $l_p$, $d_b$ (see Sec. \ref{sec:WLC-SI}) are known, the equation can be applied to infer the number of monomers $n$ in the folded structure, as is done in the Bayesian hierarchical model presented in Sec. \ref{subsec:Bayesian-clustering}. By determining $n$, we  can also distinguish whether the RNA has folded into the native state or a misfolded state. Second, assuming $n$ to be known, the equation can be used to determine the value of the WLC parameters $l_p$, $d_p$ with a least squares fitting method. For H1L12A, the native state has $n=52$, permitting us to derive the ssRNA elastic parameters.


\section{Bayesian clustering}
\label{subsec:Bayesian-clustering}

To model the unzipping experiments of RNA at low temperatures, we used a Bayesian network approach (mixture hierarchical Bayesian model). This has two advantages. First, using latent state variables in the model gives posterior distributions for the state of each data point, allowing a probabilistic soft clustering of each unfolding trace, i.e. the probability of the RNA being misfolded or native is assigned to each point. Second, using appropriate likelihood functions in the model gives a range of useful physical parameters, such as the mode and scale parameters of the rupture force distribution of each state. These parameters are related to the force average and variance. The latter gives us estimates of the distance to the transition state, $x^{\ddag}$, (see Sec. \ref{subsec:BE}), and the weight of each state, native and misfolded, in the total population.

We recall that Bayesian network models posit that the prior distributions of the parameters to be estimated are known. Similarly, the likelihood function to observe each data point \textit{given} these prior parameters is known. The estimation of the model parameters is then obtained by computing the posterior distribution of the model, given by the Bayes theorem:
\begin{equation} 
    \label{eq:bayes-high-level}
    Posterior \propto \color{blue} Likelihood \color{black} \times \color{teal} Prior \color{black} \, , 
\end{equation} 
which is often done in practice with Monte Carlo methods. 

In RNA unzipping experiments, the model data points are the pairs ($f$, $x_r$) that characterize the rupture force and released extension of each unfolding event in the forward unzipping process. The model core idea is that the extension ($x_r$) released in an unfolding event depends both on the initial folded state of the molecule (through the number of released monomers $n$, see Sec. \ref{subsec:number-nt}) and the rupture force, $f$, since force distributions are state-dependent, Sec. \ref{subsec:BE}. In practice, we use Eq.\eqref{eq:bayes-precursor} assuming that it is valid down to the presence of experimental noise, characterized as the difference between the r.h.s and l.h.s of Eq.\eqref{eq:bayes-precursor} and which we posit to be Laplace distributed around $0$ with precision $t$ (we comment further on this distribution choice at the end of the section). Therefore, for each data point ($f_i$, $x_{r,i}$), we have:
\begin{equation}\label{eq:laplace}
    x_{r,i} - n_{z_i} \cdot d_B \cdot f_{\rm WLC}^{-1}(f_i) \sim \mathrm{Laplace}(0, t) \, ,
\end{equation}
where the dependence on the number of monomers released in an unfolding event, $n$, is introduced through the use of the so-called latent variable $z_i$, which captures the initial state of a trajectory for each data point $i=1,.., N$. Here, we use the shorthand notation $z_i = 1$ for the native state and $z_i = 2$ for the misfolded state. 

The second core idea of the Bayesian classification consists of explicitly modeling the state dependency of the rupture force distribution. We posit that the parameters underpinning the rupture force distribution depend on the latent variable $z_i$. More specifically, we assume that rupture forces are Gompertz distributed with mode $M$ and scale $1/s$, and we have therefore set $M\equiv M_{z_i}$ and $s\equiv s_{z_i}$ with different values for the native/misfolded states. The overall likelihood of observing an experimental point $(f_i, x_{r, i})$ is obtained by putting all these elements together: 
\begin{equation}
    \color{blue} Likelihood \color{black} =  p \left( \overline{f_{\rm WLC}^{-1}}(f_{i}, n_{z_i})  - x_{r, i} \vert \, 0,  t \right) \times p\left(f_i \vert \, M_{z_i}, s_{z_i} \right) \times p\left(z_i \vert \, \vec{w}\right)  \, .
\end{equation}
The first term on the r.h.s is based on Eq.\eqref{eq:laplace} as described above, with the shortened notation $ \overline{f_{\rm WLC}^{-1}}(f_{i}, n_{z_i}) \equiv n_{z_i} \cdot d_B \cdot f_{\rm WLC}^{-1}(f_i)$. The second term is given by the Gompertz likelihood mentioned above. The third term, $p\left(z_i \vert \vec{w}\right)$ represents the likelihood of the latent variable $z_i$ given a weight vector $\vec{w}=(w_1, w_2)$ whose components give the average occupancy of each state. We use for $z_i$ the standard conjugate pair of a Categorical distribution for the likelihood $p$ combined with a Dirichlet prior for $\vec{w}$. 

The formal specification of the model can then be finally completed by defining appropriate priors for the parameters we want to infer, namely $n_1$, $n_2$, $M_{1}$, $M_2$, $s_{1}$, $s_2$, $t$, and $\vec{w}$. As already mentioned, we use for $\vec{w}$ a Dirichlet prior and parameterize both $t$ and $s_{1}$,$s_2$ with gamma priors. Finally, we take normal priors for $n_{1}$,$n_2$, and Laplace priors for $M_{1}$,$M_2$. The overall prior is then given by
\begin{equation}
    \begin{split}
        \color{teal} Prior \color{black} = & p(n_{z_i} \vert \, \mu_{z_i}, \nu_{z_i} ) \times p\left(M_{z_i} \vert \, \tilde{\mu}_{z_i}, \tilde{\tau}_{z_i}\right) \times  p\left(s_{z_i} \vert \, \tilde{\phi}_{z_i}, \tilde{\omega}_{z_i}\right) \times \\ & \times p(\vec{w} \vert \, \vec{\alpha}) \times p(t \vert \, \phi, \omega) \, , 
    \end{split}
\end{equation}
where the model hyper-parameters are made explicit with $z_i=1,2$. Hyper-parameters are given by the Greek variables $\mu_{1}$, $\mu_2$, $\nu_{1}$, $\nu_2$, $\tilde{\mu}_{1}$, $\tilde{\mu}_{2}$, $\tilde{\tau}_{1}$, $\tilde{\tau}_{2}$, $\tilde{\phi}_{1}$, $\tilde{\phi}_{2}$, $\tilde{\omega}_{1}$, $\tilde{\omega}_{2}$, $\alpha$, $\phi$ and $\omega$. We emphasize that while different valid choices of priors could be made, all the priors chosen here were purposefully parameterized to be very flat in order to minimally constrain the posterior space. 

Given the likelihood function and our choice of priors, we use Bayes theorem to compute the posterior distribution of the parameters we want to infer:
\begin{equation}
    p(\{z_i\}_{i=1}^{N}; n_{1,2}; w_{1,2}; M_{1,2}; s_{1,2}; \sigma)  \propto \color{blue} Likelihood \color{black} \times \color{teal} Prior \color{black}  \, ,
\end{equation}
where we defined for convenience $\sigma = 1/t$, the inverse of the precision $t$. 
The model with all its priors, likelihood, and variables is schematically summarized in \tcb{Fig. \ref{fig:BayesianNetwork}, Supp. Info}.

We used the R library RJAGS \cite{rjagsref} to set up the Bayesian network. Posterior distributions were obtained by running at least three Monte Carlo Markov Chains (MCMC) using the RJAGS library, with a burn-in of 1000 iterations, followed by 5000 iterations. We ran the usual convergence and diagnostics test for MCMCs (Gelmann, chain intercorrelation coefficient) and visually inspected the MCMC noise term to confirm that our simulations converged. We always took the median of the posterior distribution of interest for point estimates (e.g., $n_1$, $n_2$). We  give some additional details on other important aspects of the fitting procedure: 
\begin{itemize}
    \item The rupture forces are modeled as Gompertz-distributed. This is usually a good approximation in practice and even true in the BE model, Sec. \ref{subsec:BE}. Each rupture force distribution (misfolded/native) is then parametrized by a different mode $M_{z_i}$ and scale parameter $1/s_{z_i}$. Note, however, that JAGS/RJAGS does not offer a Gompertz likelihood function by default. Therefore, we need to input the likelihood manually, using Eq.\eqref{eq:reparam-BE-F} and the zero trick.
    \item When designing the model, we initially modeled the noise term in the l.h.s of Eq.\eqref{eq:laplace} with a more standard Gaussian likelihood. However, we quickly realized that some experimental points could feature large deviations between $x_i$ and $\overline{f_{\rm WLC}^{-1}}(f_i, n_{z_i})$, deviations which skew/bias the model when assuming normality and lead to overall poor convergence performance of the Monte Carlo Markov Chain (MCMC) simulation. Hence, we choose to use a more robust Laplace likelihood, which is more accommodating when a few large outliers are present. This considerably improved the model's stability. Moreover, the Deviance Information Criterion (DIC) score of the model with Laplace likelihood was lower than with a Gaussian model, giving further confidence in this choice.
\end{itemize}


\section{The H1L12A free-energies} 
\label{sec:H1L12Aenergies}

Mechanical work measurements were extracted from unzipping data as described in \tcb{Sec. \ref{subsec:free_energy}, Methods}. The inset of \tcb{Fig. \ref{fig:H1L12AFT}A} illustrates the work measured between two fixed positions (vertical lines) for the unfolding (red) and refolding (blue) FDCs in a given $N\rightleftharpoons U$ cycle. Let $P_{\rightarrow}(W), P_{\leftarrow}(-W)$ and $\Delta G$ denote the work distributions and free energy difference between N or M and U. In \tcb{Fig. \ref{fig:H1L12AFT}A} we show $P_{\rightarrow}(W)$ and $P_{\leftarrow}(-W)$ for H1L12A above room temperature, where only N is observed. For the work, $W$, we have subtracted the energy contributions of stretching the ssRNA, the hybrid DNA/RNA handles, and the bead in the trap (Sec. \ref{subsec:free_energy}, Methods).
The free energy at zero force, $\Delta G_0$, has been obtained by applying statistical approaches such as the Bennett acceptance ratio (BAR) method \cite{rissone2022stem}. Additionally, we have also determined $\Delta G_0$ using a diffusive kinetics model for the unfolding reaction, the so-called continuous effective barrier analysis (CEBA) \cite{rico2022force} (see \tcb{Sec. \ref{sec:CEBA}}). In \tcb{Fig. \ref{fig:H1L12AdG} (right panel)}, we show $\Delta G_0$ values obtained with BAR (blue circles) and CEBA (red circles) above $25^{\circ}$C (\tcb{Table S3}) finding compatible results. 
The value of $\Delta G_0$ agrees with the Mfold prediction \cite{zuker2003mfold} at $37^{\circ}$C (black triangles). However, a large discrepancy is observed for the enthalpy and entropy values if we assume $\Delta C_p=0$, suggesting a non-zero $\Delta C_p$ \tcb{(Sec. \ref{subsec:dCp0} and main text)}. 

Using the BAR method, we have also determined $\Delta G_0$ for N and M at $7^{\circ}$C. In \tcb{Fig. \ref{fig:H1L12AFT}B}, we show work distributions for N (top) and M (bottom) along with $\Delta G_0$ estimates (grey bands), finding $\Delta G_0^N=68(16) \, k_{\rm B}T$ (38(9) kcal/mol) and $\Delta G_0^M=54(18)  \, k_{\rm B}T$ (30(10) kcal/mol) in 4mM \ce{MgCl2}. We have also measured $\Delta G_0^M$ at 1M \ce{NaCl} and extrapolated it to 400mM \ce{NaCl}, the equivalent concentration to 4mM \ce{MgCl2} according to the 100:1 salt rule \cite{rissone2022stem}. We obtain $\Delta G_0^N=37(3)$ kcal/mol and $\Delta G_0^M=31(8)$ kcal/mol in 400mM \ce{NaCl} in agreement with the magnesium data. 
Figure \tcb{Fig. \ref{fig:H1L12AdG} (left panel)} shows $\Delta G_0$ for 4mM \ce{MgCl2} (filled boxes) and 400mM \ce{NaCl} (empty boxes). These values agree with a linear extrapolation from high temperatures to $7^{\circ}$C (blue and red lines, right panel). In contrast, the Mfold prediction ($\Delta G_0^N=47$ kcal/mol, black dashed line) overestimates $\Delta G_0$ by 10 kcal/mol. 


\section{Bell-Evans model} 
\label{subsec:BE}

According to the BE model \cite{bell1978models,evans1997dynamic}, the unfolding and folding kinetic rates between the folded ($F$) state and the unfolded ($U$) state, can be written as
\begin{subequations}
\label{eq:BEkinetic_rate}
\begin{align}
	\label{eq:BEkinetic_rate1}
	k_{F\rightarrow U}(f) &= k_0 \exp \left( - \frac{B_0  - fx^{\ddag}}{k_{\rm B}T} \right) \\
	\label{eq:BEkinetic_rate2}
	k_{U\rightarrow F}(f) &= k_0 \exp \left( - \frac{B_0 - \Delta G_{FU} + f(x_U - x^{\ddag})}{k_{\rm B}T} \right) \, ,
\end{align}
\end{subequations}
where $k_0$ is the pre-exponential factor, $x^{\ddag}$ ($x_U - x^{\ddag}$) are the relative distances between state $F$ ($U$) and the transition state. $\Delta G_{FU}$ is the free energy difference between states $F$ and $U$ at zero force. 
In a pulling experiment, the force is ramped linearly with time, $f = rt$, with $r$ the experimental pulling rate. The survival probability in the folded state (F) is 
\begin{equation}
	\label{eq:Fprob0}
	\frac{dP_F(f)}{df} = - \frac{k_{F\rightarrow U}(f)}{r} P_F(f) \, .
\end{equation}
By solving Eqs.\eqref{eq:BEkinetic_rate1}, \eqref{eq:BEkinetic_rate2}, and \eqref{eq:Fprob0} we get the unfolding rupture force distribution:
\begin{equation}
    \begin{split}
        p_{F \rightarrow U}(f) = -\frac{dP_F(f)}{df}=& \frac{k_0}{r}\exp\left(\frac{k_0 \Bz T}{r x^{\ddag}} \right) \exp\left( \frac{f x^{\ddag}}{\Bz T}\right)  \times \\ & \times \exp\left(-  \frac{k_0 \Bz T}{r x^{\ddag}} \exp\left(\frac{f x^{\ddag}}{\Bz T}\right)\right) \, ,
    \end{split}
\end{equation}
while for the reverse force distribution, one can similarly obtain
\begin{equation}
    \begin{split}
        p_{U \rightarrow F} (f) = & \frac{\tilde{k}_0}{r} \exp\left(-\frac{f(x_U  - x^{\ddag})}{\Bz T} \right) \times \\ & \times \exp \left( -\frac{\tilde{k}_0 \Bz T}{r {(x_U  - x^{\ddag})}} \exp \left[ - \frac{f(x_U - x^{\ddag})}{\Bz T}\right] \right) \, ,
    \end{split}
\end{equation}
where we introduced $\tilde{k}_0 := k_0 \exp(\Delta G_{FU} / \Bz T)$. We can recognize that the forward force distribution follows a Gompertz law, with an inverse scale parameter $s_U := \Bz T / x^{\ddag}$ and with a mode given by $\mu_U = s_U \ln ( r/k_0 s_U)$. This leads to the following useful re-parametrization:
\begin{equation}\label{eq:reparam-BE-F}
    p_{F \rightarrow U} (f) = \frac{1}{s_U} \exp \left(\frac{f-\mu_U}{s_U} + \exp\left(-\frac{\mu_U}{s_U } \right) - \exp \left(  \frac{f - \mu_U}{s_U} \right) \right)  \, .
\end{equation}
Eq.\eqref{eq:reparam-BE-F} is very convenient as it is expressed in terms of the quantities $s_U, \mu_U$ whose order of magnitude can be easily estimated from experimental data (unlike $k_0$, $x^{\ddag}$). For this reason, we used it both for MLE estimation (to retrieve $s_U$ and then $x^{\ddag}$) and as a likelihood function in our Bayesian clustering algorithm (see Sec. \ref{subsec:Bayesian-clustering}).

The distance to the transition state $x^{\ddag}$ is often expressed as a function of the variance of the rupture force distribution. Notice that there is no simple closed formula for $Var(X)$ when $X$ is a Gompertz-distributed random variable. For Gompertz distributions, the following approximation can be derived, $Var(X) \cong s_{U}^{2}\frac{\pi^2}{6}$ \cite{lenart2012gompertz}. It holds to a very good accuracy for the rupture force distributions measured in pulling experiments. As $\sqrt{\pi^2 / 6}  \approx 1 $, the rupture force distribution standard deviation is approximately equal to the inverse distance to the transition state per unit of $\Bz T$ in the BE model.


\section{Continuous Effective Barrier Analysis}
\label{sec:CEBA}

In the BE model, the height of the kinetic barrier decreases linearly with the applied force, $B(f) = B_0 - f x^{\ddag}$. This hypothesis is relaxed in the kinetic diffusion (KD) model, which assumes the folding reaction as a diffusive process in a one-dimensional force-dependent free-energy landscape. 
The Continuous Effective Barrier Approach (CEBA) is based on the KD model. It can be used to extract the force-dependent behavior of the kinetic barrier from unzipping experiments \cite{alemany2016mechanical,alemany2017force}. In CEBA, the effective barrier between the folded ($F$) and the unfolded state ($U$), $B(f)$, is derived by imposing the detailed balance condition between the unfolding, $k_{FU}(f)$, and folding, $k_{FU}(f)$, kinetic rates (see Eqs.\eqref{eq:BEkinetic_rate}):
\begin{subequations}
\label{eq:CEBAkinetic_rate}
\begin{align}
	\label{eq:CEBAkinetic_rate1}
	k_{F\rightarrow U}(f) &= k_0 \exp \left( - \frac{B(f)}{\Bz T} \right) \\
	\label{eq:CEBAkinetic_rate2}
	k_{U\rightarrow F}(f) &= k_{F\rightarrow U}(f) \exp \left( \frac{\Delta G_{FU}(f)}{\Bz T} \right) \, ,
\end{align}
\end{subequations}
where $k_0$ is the attempt rate, $B(f)$ is the effective barrier at force $f$, and $\Delta G_{FU}(f)=G_U(f)-G_F(f)$ is the folding free energy at force $f$. The latter term is given by
\begin{equation}
	\label{eq:CEBAdG}
	\Delta G_{FU}(f) = \Delta G_{0} - \int_0^f \left(x_U(f^{\prime}) - x_F(f^{\prime}) \right) df^{\prime} \, ,
\end{equation}
where $\Delta G_{0}$ is the folding free energy between $F$ and $U$ at zero force, and the integral accounts for the free energy change upon stretching the molecule in state $U$ ($F$) at force $f$.

We can derive two estimates for $B(f)$ by computing the logarithms of Eqs.\eqref{eq:CEBAkinetic_rate1} and \eqref{eq:CEBAkinetic_rate2}, which give
\begin{subequations}
\label{eq:CEBAbarrier}
\begin{align}
	\label{eq:CEBAbarrier1}
	\frac{B(f)}{\Bz T} &= \log{k_0} - \log{k_{F\rightarrow U}(f)} \\
	\label{eq:CEBAbarrier2}
	\frac{B(f)}{\Bz T} &= \log{k_0} - \log{k_{U\rightarrow F}(f)} + \frac{\Delta G_{FU}(f)}{\Bz T} \, .
\end{align}
\end{subequations}
By imposing the continuity of the two estimations of $B(f)$ in Eqs.\eqref{eq:CEBAbarrier}, we can measure the folding free energy at force $f$, $\Delta G_{FU}(f)$. The free energy of the stretching contribution in $\Delta G_{FU}(f)$ Eq.\eqref{eq:CEBAdG} can be measured from the unfolded branch. Matching Eqs.\ref{eq:CEBAbarrier1} and \ref{eq:CEBAbarrier2} permits us to directly estimate the folding free energy at zero force, $\Delta G_{0}$. Details can be found in \cite{alemany2017force, rico2022force}.


\section{H1L12A thermodynamics for $\Delta C_p=0$}
\label{subsec:dCp0}

A fit to the linear function $\Delta G_0 = \Delta H_0 -T\Delta S_0$ gives estimates for the folding enthalpy ($\Delta H_0$) and entropy ($\Delta S_0$) by assuming $\Delta C_p=0$. We get $\Delta H^{\rm BAR}_0 = 110(30) \, \rm kcal \, mol^{-1}$, $\Delta S^{\rm BAR}_0 = 240(10) \, \rm cal \, mol^{-1}K^{-1}$ (continuous blue line in Fig. \ref{fig:H1L12AdG}) and $\Delta H^{\rm CEBA}_0 = 100(30) \, \rm kcal \, mol^{-1}$, $\Delta S^{\rm CEBA}_0 = 230(8) \, \rm cal \, mol^{-1}K^{-1}$ (continuous red line in Fig. \ref{fig:H1L12AdG}). Our results for $\Delta G_0$ agree with the Mfold prediction at $37^{\circ}$C (black triangles in Fig. \ref{fig:H1L12AdG}) above room temperature. However, a discrepancy is observed for the enthalpy and entropy values, $\Delta H^{\rm Mfold}_0 = 196 \, \rm kcal \, mol^{-1}$, $\Delta S^{\rm Mfold}_0 = 533 \, \rm cal \, mol^{-1}K^{-1}$, almost twice our numbers. 
In contrast, by assuming $\Delta C_p \neq 0$ and applying the Clausius-Clapeyron equation (see main text), we measured $\Delta H^{\rm BAR}_0 = 163(6) \, \rm kcal \, mol^{-1}$ and $\Delta S^{\rm BAR}_0 = 420(20) \, \rm cal \, mol^{-1}K^{-1}$ at $37^{\circ}$C. In this case, we only report results obtained from $\Delta G_0$ measurements derived with the BAR method, available at all $T$ for N and at $7{\deg}$C for M.


\section{H1L4A and H1L12A thermodynamics for $\Delta C_p\neq0$}
\label{subsec:dCp}

We have measured the enthalpy $\Delta H_0$ and entropy $\Delta S_0$ of N at all temperatures for hairpins H1L4A and H1L12A using the Clausius-Clapeyron equation under an applied force \cite{rico2022molten}. Both $\Delta H_0$ and $\Delta S_0$ are temperature dependent indicating a finite $\Delta C_p=\frac{\partial \Delta H}{\partial T}=T\frac{\partial \Delta S}{\partial T}$. We observe two distinct regimes, above (hot, H) and below (cold, C) $\sim 20^{\circ}$C (see \tcb{Fig. 5B, main text}). By separately fitting the two regimes to the function $\Delta S_0(T)=\Delta S_{\rm H(C)}+\Delta C_p\log(T/T_{\rm H(C)})$ where $\Delta S_{\rm H(C)}$ is the entropy at the hot (cold) denaturation temperature $T_{\rm H(C}$, we find $\Delta C_p^{\rm H} = 1.5(2) \cdot 10^3$ cal mol$^{-1}$K$^{-1}$ and $\Delta C_p^{\rm C} = 5.8(4) \cdot 10^3$ cal mol$^{-1}$K$^{-1}$ for H1L4A, and $\Delta C_p^{\rm H} = 1.5(2) \cdot 10^3$ cal mol$^{-1}$K$^{-1}$ and $\Delta C_p^{\rm C} = 8(1) \cdot 10^3$ cal mol$^{-1}$K$^{-1}$ for H1L12A. 
Notice that by measuring $\Delta C_p^{\rm H(C)}$ from the enthalpy $T$-dependence, we obtain compatible results: $\Delta C_p^{\rm H} = 1.3(2) \cdot 10^3$ cal mol$^{-1}$K$^{-1}$ and $\Delta C_p^{\rm C} = 6.0(4) \cdot 10^3$ cal mol$^{-1}$K$^{-1}$ for H1L4A, and $\Delta C_p^{\rm H} = 1.7(3) \cdot 10^3$ cal mol$^{-1}$K$^{-1}$ and $\Delta C_p^{\rm C} = 7(1) \cdot 10^3$ cal mol$^{-1}$K$^{-1}$ for H1L12A.

$\Delta C_p$ values permit us to determine hot and cold denaturation temperatures, which are $T_{\rm H} = 377(3)$K and $T_{\rm C} = 222(3)$K for H1L4A, and $T_{\rm H} = 366(4)$K and $T_{\rm C} = 228(4)$K for H1L12A.
Despite the differences in $\Delta C_p^{\rm C}$ between H1L4A and H1L12A, in both cases, stability is maximum at $T_S \sim 5^{\circ}$C where $\Delta S_0=0$. Moreover, cold denaturation is predicted at $T_C \sim -50(4)^{\circ}$C for both sequences. This concurrency suggests that maximum stability and cold denaturation temperatures are sequence-independent.

Finally, we report the entropy at the hot and cold denaturation temperatures from the fit $\Delta S_{\rm H} = 630 (40)$ cal mol$^{-1}$K$^{-1}$ and $\Delta S_{\rm C} = -1340 (90)$ cal mol$^{-1}$K$^{-1}$ for H1L4A, and $\Delta S_{\rm H} = 670 (40)$ cal mol$^{-1}$K$^{-1}$ and $\Delta S_{\rm C} = -1600 (200)$ cal mol$^{-1}$K$^{-1}$ for H1L12A.

\newpage
\clearpage


\renewcommand{\thesection}{Supplementary Figures}
\section{}

\null
\vfill
\begin{figure*}[h]
    \centering
    \includegraphics[width=\textwidth]{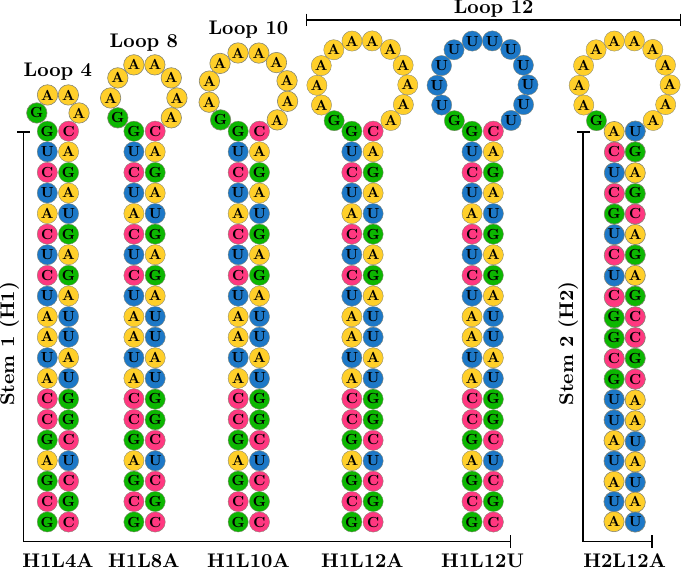}
    \caption{\textbf{Sequence of the studied RNA hairpins.} Each hairpin is named after the stem sequence (H1 or H2), the loop size (L4, L8,  L10, or L12), and the loop composition (poly-A or poly-U). The sequences are reported in Table \ref{tab:RNAhairpins}}
    \label{fig:Seq}
\end{figure*}
\vfill
\clearpage
\newpage

\null
\vfill
\begin{figure*}[h]
    \centering
    \includegraphics[width=\textwidth]{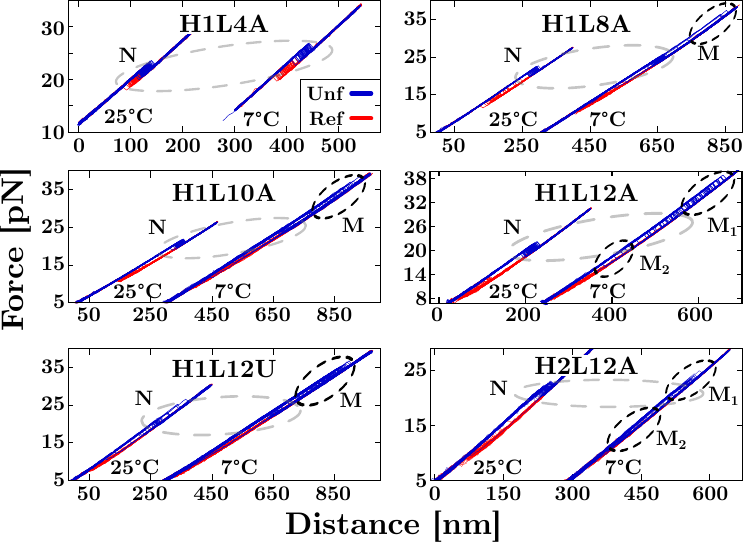}
    \caption{\textbf{RNA unzipping experiments in sodium.} Experimental FDCs ($25^{\circ}$C and $7^{\circ}$C) measured at 1M NaCl for all the studied hairpins. At low-$T$, all molecules exhibit analogous behavior to that observed at 4mM MgCl$_2$.}
    \label{fig:FDCsodium}
\end{figure*}

\null
\vfill
\begin{figure*}[h]
    \centering
        \includegraphics[width = \textwidth]{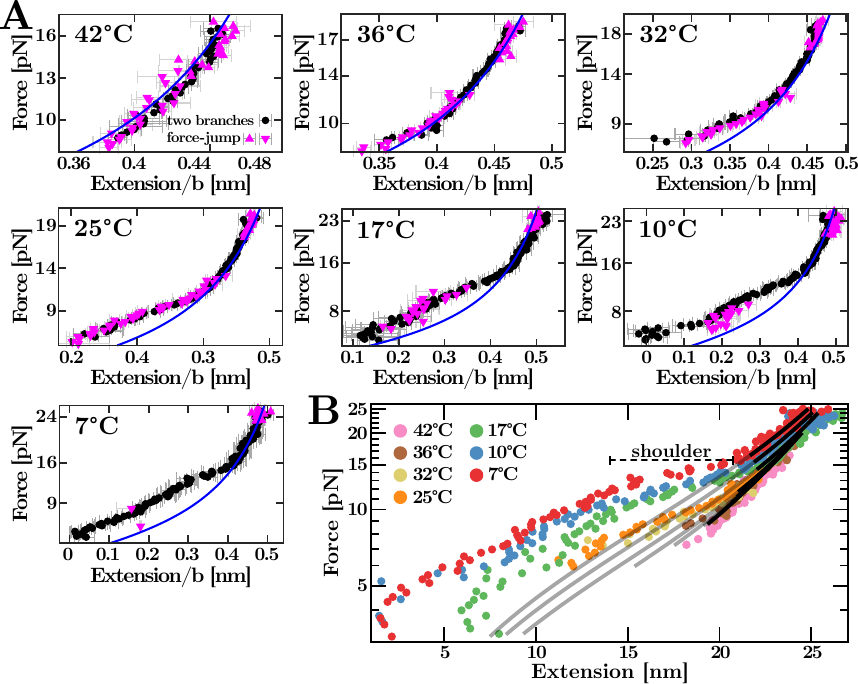}
        \caption{\textbf{$T$-dependent H1L12A ssRNA elastic response.} (\textbf{A}) Force versus the ssRNA extension per base at all the studied temperatures. Two methods have been used to extract the molecular extension of the ssRNA: the force-jump (magenta triangles up -unfolding- and down -refolding-) and the two-branches method (black circles) \cite{zhang2009mechanoenzymatic,alemany2014determination}. Blue lines are the fits to the WLC.(\textbf{B}) Overview of the $T$-dependent ssRNA elastic response. Only data above the shoulder (horizontal-dashed segment) have been used for the fit (black lines). Grey lines extrapolate the WLC fitting curves to the non-fitting regions.} 
        \label{fig:H1L12ssRNA}
\end{figure*}
\vfill
\clearpage
\newpage

\null
\vfill
\begin{figure*}[h]
    \centering
    \includegraphics[width=0.9\textwidth]{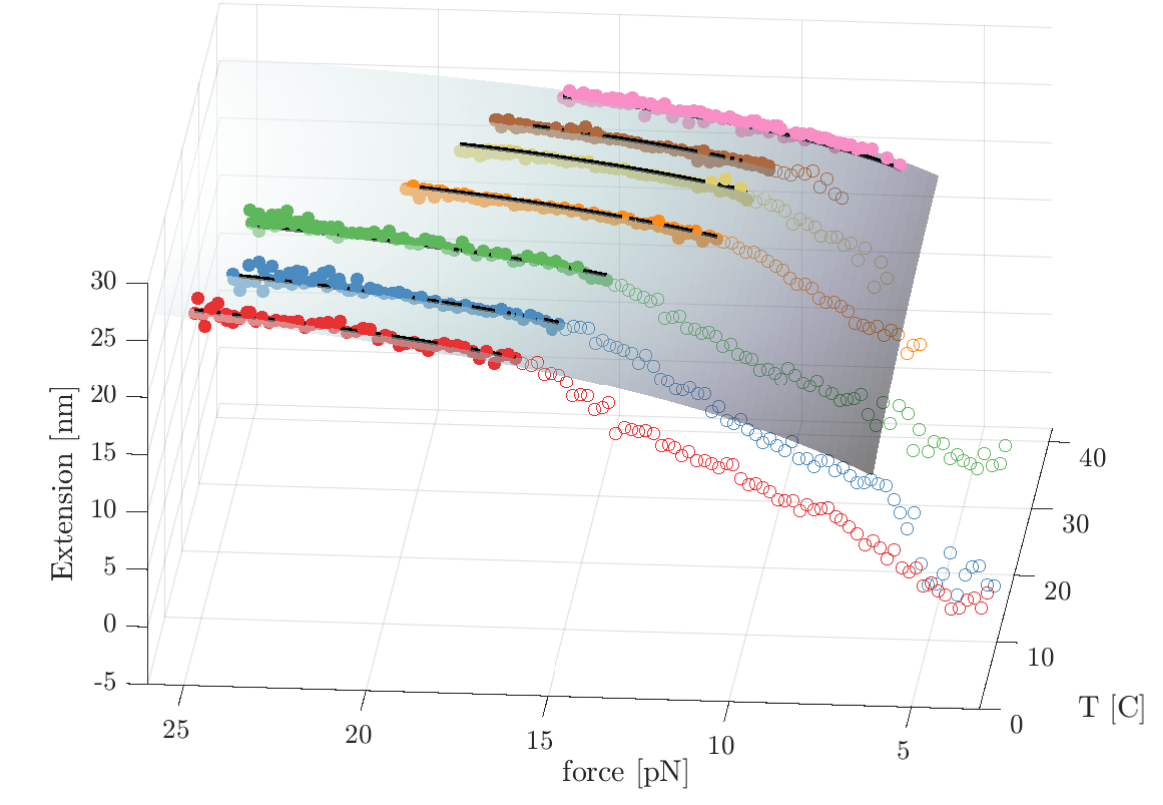}
        \caption{\textbf{Multi-$T$ fit of the H1L12A ssRNA elastic response}. We simultaneously fit the relation $l_p = l_p(T) = a_1 T + b_1$, $l_B = l_B(T) = a_2 T + b_2$ on data points at all temperatures. This gives for $l_p$ the values $a_1 = 0.135 \pm 0.006$ \r{A}/C, $b_1=3.5 \pm 0.1$ \r{A} and for $l_B$ the values $ a_2 = -0.023 \pm 0.002$ \r{A}/C, $b_2 = 7.37 \pm 0.05$ \r{A}. The fit is performed over the filled symbols only. The three-dimensional $T$-force-extension surface is represented in light grey. The black lines plot force-extension cross-sections at a given temperature (red, $T=7\deg$C; blue, $T=10\deg$C; green, $T=17\deg$C; orange, $T=25\deg$C; yellow, $T=32\deg$C; brown, $T=36\deg$C; pink, $T=42\deg$C).}
        \label{fig:ElasticMultiFit}
\end{figure*}
\vfill
\clearpage
\newpage

\vfill
\begin{figure*}[h]
    \centering
    \includegraphics[width=\textwidth]{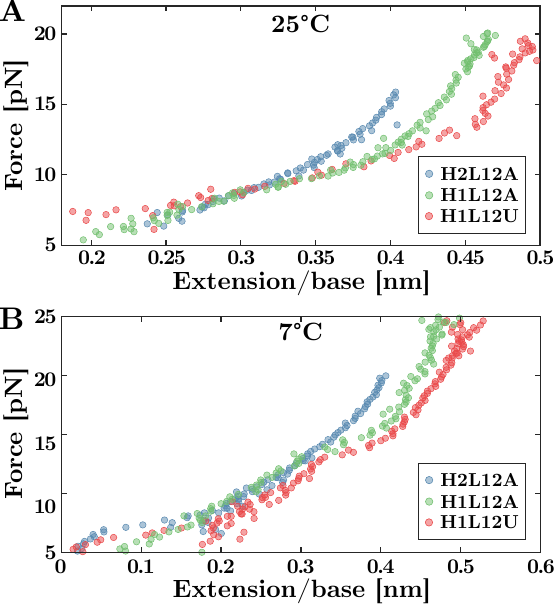}
    \caption{\textbf{$T$-dependence of the H1L12A, H2L12A, and H1L12U ssRNA elastic response.} Results are shown at $T=25^{\circ}$C (panel A) and $7^{\circ}$C (panel B).. Extension is normalized per base by dividing the measured extension by each hairpin's total number of bases. The normalized extensions do not collapse as different sequences feature different elastic properties (see main text).}
    \label{fig:ElasticOthers} 
\end{figure*}
\vfill
\clearpage
\newpage

\vfill
\begin{figure*}[h]
    \centering
    \includegraphics[width=\textwidth]{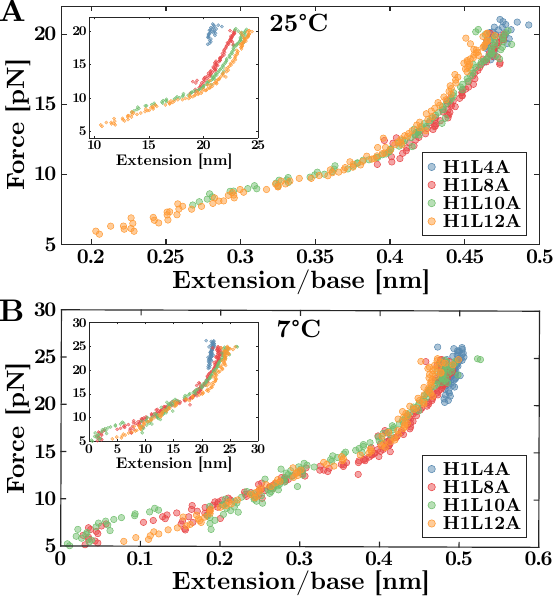}
    \caption{\textbf{$T$-dependent ssRNA response for H1L4A, H1L8A, H1L10A, and H1L12A.} Results are shown at $T=25^{\circ}$C (panel A) and $7^{\circ}$C (panel B). If we plot the force versus the total extension, data for different hairpins do not collapse (insets of A and B). In contrast, upon normalizing the extension per base, the force-extension curves of all hairpins collapse into a master curve (main A and B). Extension is normalized per base by dividing the measured extension by each hairpin's total number of bases.}
    \label{fig:ElasticCD4Family} 
\end{figure*}
\vfill
\clearpage
\newpage

\null
\vfill
\begin{figure*}[h]
    \centering
        \includegraphics[width = \textwidth]{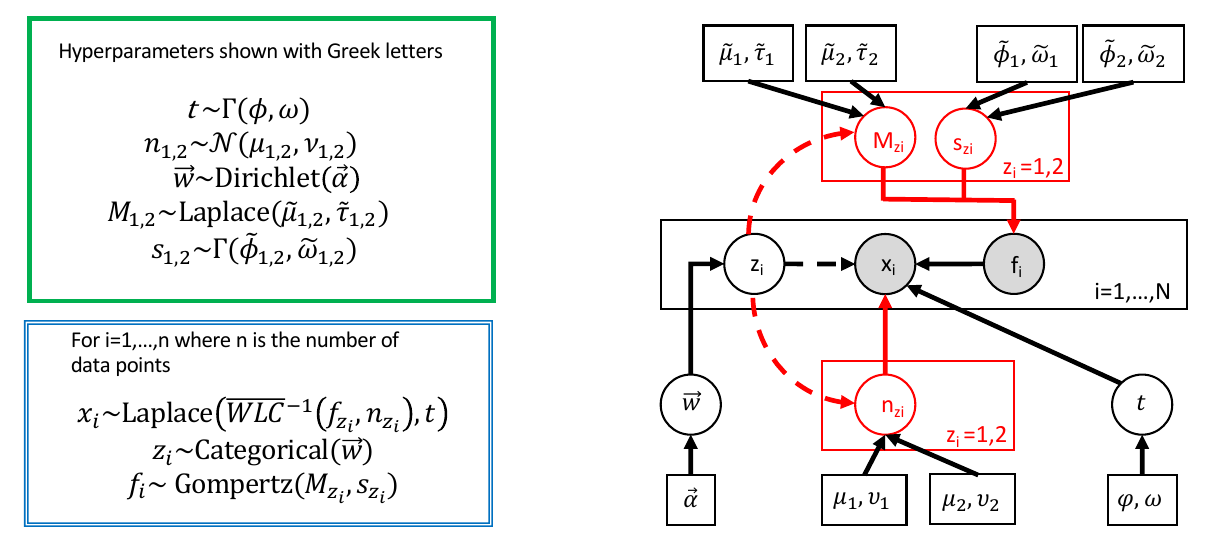}
        \caption{\textbf{The Bayesian classification algorithm.} (\textbf{Left}) Specification of the prior (in green) and likelihood (in blue) functions used. The hyper-parameters used are indicated by Greek letters. $\Gamma$ stands for the gamma distribution, $\mathcal{N}$ for the normal distribution. (\textbf{Right}) Probabilistic graph view of the Bayesian network used. Misfolded and native states are represented by the superscript $1,2$ and are encoded in the latent variables $z_i = 1,2$. We highlight in red the part of the model that depends on $z_i$: the rupture force distribution through its mode $M_{z_i}$ and scale $1/s_{z_i}$ parameters and the number of monomers released within an unfolding event $n_{z_i}$.} 
        \label{fig:BayesianNetwork}
\end{figure*}
\vfill
\clearpage
\newpage

\null
\vfill
\begin{figure*}[h]
    \centering
        \includegraphics[width = \textwidth]{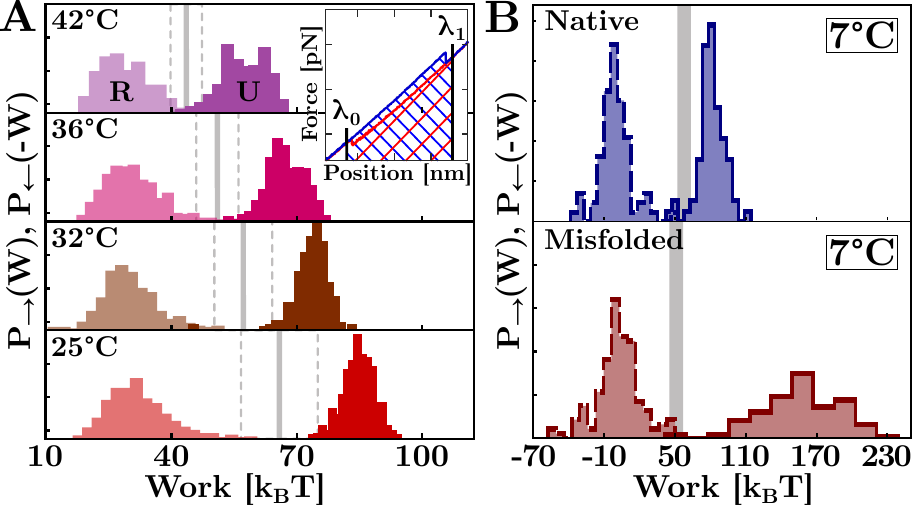}
        \caption{\textbf{Fluctuation theorem applied on the H1L12A.} (\textbf{A}) Unfolding (U) and refolding (R) work distributions, and $\Delta G_0$ estimates (gray thick lines) at 4mM \ce{MgCl2} above $25^{\circ}$C. Results are averages over 5-6 molecules. Dashed grey lines show the hysteresis region. \textbf{Inset.} The measured work equals the area under the FDC for unfolding (blue) and refolding (red). (\textbf{B}) $P_{\rightarrow}(W)$ (solid line) and $P_{\leftarrow}(-W)$ (dashed line) for N (blue) and M (red) states at 4mM \ce{MgCl2} and $7^{\circ}$C.} 
        \label{fig:H1L12AFT}
\end{figure*}
\vfill
\clearpage
\newpage

\null
\vfill
\begin{figure*}[h]
    \centering
        \includegraphics[width = 0.8\textwidth]{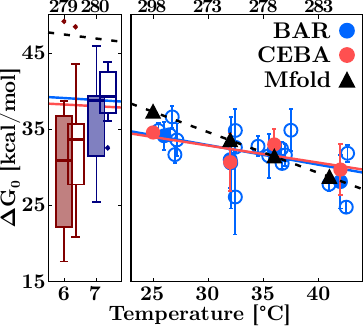}
        \caption{\textbf{$T$-dependence of the H1L12A free energy for N and M}. $\Delta G_0$ values at $7^{\circ}$C in 4mM \ce{MgCl2} (solid boxes) and 400mM \ce{NaCl} (empty boxes) \textbf{(left)}, and above $25^{\circ}$C at 4mM \ce{MgCl2} for N \textbf{(right)}. \textbf{(Left)} Results are shown with a box-and-whisker plot indicating the data median (horizontal thick line), first and third quartiles (box), 10th and 90th percentiles (whiskers), and outliers (dots). \textbf{(Right)} BAR (blue) and CEBA (red) results are compared with Mfold prediction (black). For BAR, we show results for different molecules (empty circles) and their averages (solid circles). Temperature axis in $^{\circ}$C (bottom label) and K (top label).} 
        \label{fig:H1L12AdG}
\end{figure*}
\vfill
\clearpage
\newpage

\null
\vfill
\begin{figure*}[h]
    \centering
        \includegraphics[width = 0.8\textwidth]{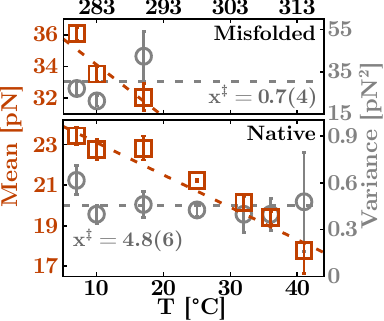}
        \caption{\textbf{Transition state distance in H1L12A.} Mean (orange squares) and variance (grey circles) of the rupture force distributions for native (bottom) and misfolded (top) at each experimental $T$ (in Celsius and Kelvin) for H1L12A. The dashed grey lines denote the average variance used to measure the transition state distances for N and M.} 
        \label{fig:TSdistance}
\end{figure*}
\vfill
\clearpage
\newpage

\clearpage


\renewcommand{\thesection}{Supplementary Tables}
\section{}

\vfill
\begin{table*}[h]
\centering
    \begin{tabular}{r|ll}
        \toprule
        \textbf{Molecule} &  \textbf{Stem Sequence} &  \textbf{Loop Sequence}\\
        \midrule
        $\rm H1L4A$ & GCGAGCCAUAAUCUCAUCUG & GAAA \\
        $\rm H1L8A$ & GCGAGCCAUAAUCUCAUCUG & GAAAAAAA \\ 
        $\rm H1L10A$ & GCGAGCCAUAAUCUCAUCUG & GAAAAAAAAA \\ 
        $\rm H1L12A$ & GCGAGCCAUAAUCUCAUCUG & GAAAAAAAAAAA \\ 
        $\rm H1L12U$ & GCGAGCCAUAAUCUCAUCUG & GUUUUUUUUUUU \\ 
        $\rm H2L12A$ & AUAUAUUGCGGCUCUGCUCA & GAAAAAAAAAAA\\
        \bottomrule
    \end{tabular}
    \caption{Sequences of the six RNA hairpins ($5^{\prime} \rightarrow 3^{\prime}$, from left to right).} \label{tab:RNAhairpins}
\end{table*}
\vfill
\clearpage
\newpage

\null
\vfill
\begin{table*}[h]
\centering
\label{tab:H1l12Aelastic}
    \begin{tabular}{cc|cc|cc|cc}
        \multicolumn{2}{c}{ } & \multicolumn{2}{c}{H1L12A}  & \multicolumn{2}{c}{H1L12U}  & \multicolumn{2}{c}{H2L12A} \\
        \toprule
        \textbf{T [$^\circ$C]} & \textbf{T [K]} & $\mathbf{l_p}$ & $\mathbf{d_b}$ & $\mathbf{l_p}$&$\mathbf{d_b}$ & $\mathbf{l_p}$&$\mathbf{d_b}$\\
        \midrule
        $42$ & $315$ & $0.99(3)$ & $0.64(1)$ & & \\
        $36$ & $309$ & $0.81(2)$ & $0.66(1)$ & & \\
        $32$ & $305$ & $0.79(3)$ & $0.65(1)$ & & \\
        $25$ & $298$ & $0.70(2)$ & $0.67(1)$ & $0.60(3)$ & $0.75(1)$ & $0.55(3)$ & $0.67(2)$\\
        $17$ & $290$ & $0.46(2)$ & $0.76(1)$ & & \\
        $10$ & $283$ & $0.42(3)$ & $0.75(2)$ & & \\
        $7$ & $280$ & $0.38(3)$ & $0.75(2)$ & $0.35(3)$ & $0.82(2)$ & $0.28(2)$ & $0.77(2)$ \\
    \bottomrule
\end{tabular}
\caption{Temperature dependence of the persistence length ($l_p$) [nm] and the interphosphate distance ($d_b$) [nm] of the different dodecaloop hairpins. The error (in brackets) refers to the last digit.}
\end{table*}
\vfill
\clearpage
\newpage

\null
\vfill
\begin{table*}[h]
\centering
\label{tab:H1L12AdG_highT}
    \begin{tabular}{cc|ccc}
        \toprule
        \textbf{T [$^\circ$C]} & \textbf{T [K]} & \textbf{BAR} & \textbf{CEBA} & \textbf{Mfold}\\
        \midrule
        $42$ & $315$ & $28$ $(3)$ & $30$ $(3)$ & $28$\\
        $36$ & $309$ & $32$ $(1)$ & $33$ $(2)$ & $31$\\
        $32$ & $305$ & $31$ $(4)$ & $31$ $(4)$ & $34$\\
        $25$ & $298$ & $34$ $(2)$ & $35$ $(1)$ & $37$\\
        \bottomrule
    \end{tabular}
    \caption{$\Delta G_0$ [kcal/mol] values of N for $\rm H1L12A$ in the high-$T$ regime measured with BAR and CEBA methods compared to predictions by Mfold. The error (in brackets) refers to the last digit.}
\end{table*}
\vfill
\clearpage
\newpage

\null
\vfill
\begin{table*}[h]
\centering
\label{tab:H1L12AdG_lowT} 
    \begin{tabular}{ccc|cc}
        \toprule
        \textbf{T [$^\circ$C]} & \textbf{T [K]} & \textbf{State} & \textbf{Mg$^{2+}$} & \textbf{Na$^{+}$}\\
        \midrule
        \multirow{2}{*}{$7$} & \multirow{2}{*}{$280$} & N & $38$ $(9)$ & $37$ $(3)$ \\
        & & M & $31$ $(10)$ & $31$ $(8)$ \\
        \bottomrule
    \end{tabular}
\caption{$\Delta G_0$ [kcal/mol] values of M for $\rm H1L12A$ at $T=7^{\circ}$C derived with the BAR method. To be compared, the $100/1$ equivalence rule between monovalent and divalent salt concentrations has been applied to the sodium results. The error (in brackets) refers to the last digit.}
\end{table*}
\vfill
\clearpage
\newpage

\null
\vfill
\begin{table*}[h]
\centering
\label{tab:H1l12Athermo}
    \begin{tabular}{cc|cc}
        \toprule
        \textbf{T [$^\circ$C]} & \textbf{T [K]} & $\mathbf{\Delta H \, [kcal \, mol^{-1}}]$ & $\mathbf{\Delta S \, [cal \, mol^{-1} \, K^{-1}}]$\\
        \midrule
        $42$ & $315$ & $165$ $(7)$ & $433$ $(18)$ \\
        $36$ & $309$ & $163$ $(6)$ & $421$ $(19)$ \\
        $32$ & $305$ & $152$ $(7)$ & $396$ $(18)$ \\
        $25$ & $298$ & $139$ $(5)$ & $351$ $(17)$ \\
        $17$ & $290$ & $114$ $(8)$ & $263$ $(15)$ \\
        $10$ & $283$ & $71$ $(6)$ & $110$ $(9)$ \\
        $7$ & $280$ & $39$ $(10)$ & $2$ $(3)$ \\
    \bottomrule
\end{tabular}
\caption{Temperature dependence of the enthalpy ($\Delta H$) and entropy ($\Delta S$) of N for H1L12A. The error (in brackets) refers to the last digit(s).}
\end{table*}
\vfill
\clearpage
\newpage

\null
\vfill
\begin{table*}[h]
\centering
\label{tab:H1l4Athermo}
    \begin{tabular}{cc|ccc}
        \toprule
        \multirow{2}{*}{\textbf{T [$^\circ$C]}} & \multirow{2}{*}{\textbf{T [K]}} & $\mathbf{\Delta G_0}$ & $\mathbf{\Delta H}$ & $\mathbf{\Delta S}$\\
        & & $\mathbf{[kcal \, mol^{-1}]}$ & $\mathbf{[kcal \, mol^{-1}]}$ & $\mathbf{[cal \, mol^{-1} \, K^{-1}]}$ \\
        \midrule
        40 & 313 & 30 (3) & 345 (35) & 138 (11) \\
        32 & 305 & 33 (4) & 316 (34) & 130 (11) \\
        25 & 298 & 37 (3) & 269 (33) & 117 (10) \\
        15 & 288 & 39 (5) & 155 (27) & 84 (9) \\
        10 & 283 & 38 (4) & 61 (19) & 56 (7) \\
        7 & 280 & 38 (4) & -12 (12) & 34 (5) \\
    \bottomrule
\end{tabular}
\caption{Temperature dependence of the free energy ($\Delta G_0$), enthalpy ($\Delta H$), and entropy ($\Delta S$) of N for H1L4A. The error (in brackets) refers to the last digit(s).}
\end{table*}
\vfill
\clearpage
\newpage